\documentclass[twocolumn,aps,pre,showpacs,epsfig,amsmath,amssymb,eqsecnum]{revtex4}
\usepackage{bbm}
\usepackage{mathrsfs}
\usepackage{amsfonts}
\usepackage{color}
\usepackage[dvips]{graphicx}
\newcommand{\rv}{{\vec r}}

\newcommand{\xv}{{\vec x}}

\newcommand{\be}{\begin{equation}}
\newcommand{\ee}{\end{equation}}
\newcommand{\bea}{\begin{eqnarray}}
\newcommand{\eea}{\end{eqnarray}}
\begin{document}
\title{The Poisson-Boltzmann Theory for Two Parallel Uniformly Charged Plates}
\author{Xiangjun Xing}
\address{Institute of Natural Sciences and Department of Physics, 
Shanghai Jiao Tong University,
Shanghai, 200240 China}
\email{xxing@sjtu.edu.cn}

\date{\today} 
\begin{abstract} 
We solve the nonlinear Poisson-Boltzmann equation for two parallel and likely charged plates both inside a symmetric elecrolyte, and inside a $2:1$ asymmetric electrolyte, in terms of Weierstrass elliptic functions.  From these solutions we derive the functional relation between the surface charge density, the plate separation, and the pressure between plates.  For the one plate problem, we obtain exact expressions for the electrostatic potential and for the renormalized surface charge density, both in symmetric and in asymmetric electrolytes.   For the two plate problems, we obtain new exact asymptotic results in various regimes.  
\end{abstract}

\maketitle
\section{Introduction}
\label{sec:intro}

When a charged object is inserted into an electrolyte, it attracts ions of opposite charge and repels ions of like charge.  This leads to the well-known phenomenon of screening: the total electrostatic potential, due to both the external charges and the electrolyte, is exponential damped as a function of distance from the charged object.  This phenomenon was first studied by Debye and H\"{u}ckel (DH) \cite{Debye-Huckel}, and is often called Debye screening.  At the level of mean field theory, the screened potential satisfies the so-called Poisson-Boltzmann (PB) equation, which, for a symmetric electrolyte with ion charges $\pm q$, and ion number density $n$ for each specie, is given by
\be
 - \epsilon \Delta \phi + 2 n q \, \sinh \beta q \phi = 0,
\label{PB-symmetric} 
\ee
where $\beta$ is the Boltzmann factor, $q= 1.6\times 10^{-19}C$ the charge of an electron.  It is convenient to use the dimensionless potential $\Psi = \beta q \phi$, and to measure length in terms of the Debye length 
\be
\ell_{DB}= \kappa^{-1} =  \sqrt{\epsilon/2 \beta q^2 n},
\quad {\rm q:-q.}
\label{Debye-length-def}
\ee  
If the surface charge density on the object is low so that $|\Psi |\ll 1$ everywhere, the PB Eq.~(\ref{PB-symmetric}) can be further linearized to yield:
\be
- \Delta \Psi + \Psi = 0, 
\label{PB-linear}
\ee
which admits the famous screened Coulomb potential $e^{-r}/4 \pi r$ as its Green's function.   

Linearization is however not appropriate for strongly charged objects.  On the other hand, the full PB equation is difficult to solve due to its nonlinear nature.   For the simple case of  one positively charged plate of infinite size, the exact solution has been known since the time of Verwey and Overbeek \cite{Verwey-Overbeek}: 
\be
\Psi^+_{q:-q}(z) = 2
\log \frac{1+ e^{- z }}
{1- e^{- z  }},  
\label{potential-one-plate}
\ee
where $z$ is the coordinate perpendicular to the plate.  The potential by a negatively charged plate is just the negative of Eq.~(\ref{potential-one-plate}).  
Note that Eq.~(\ref{potential-one-plate}) has a logarithmic singularity at $z = 0$.  

The general solution to the one plate problem is $\Psi(z+z_0)$ with an arbitrary parameter $z_0$.  Normally one would have to determine the constant $z_0$ with the surface charge density $\sigma$ and the position of the plate fixed.  For our purpose, however, it is much more convenient to fix the electrostatic potential to be Eq.~(\ref{potential-one-plate}) and let $z_0$ be the position of the plate, {\em which should be adjusted} according to the Neumann boundary condition 
\be
\Psi'(z_0) 
= \beta q \ell_{DB} \frac{\partial \phi(z_0)}{\partial n}
= - \frac{4}
{e^{\kappa  z_0 } - e^{- \kappa z_0 }}
= - \frac{\beta q\sigma}{\epsilon \kappa}. 
\label{BC-sigma}
\ee
Defining a Gouy-Chapman length $\ell_{GC}$ and a dimensionless surface charge density $\eta$ via
\bea
\ell_{GC} &=& \epsilon/ q \beta \sigma
\quad \quad \quad \quad \quad \mbox{Gouy-Chapman length},
\\
\eta &=& \frac{\beta q \sigma}{ \epsilon \kappa} 
=  \frac{\ell_{DB}}{\ell_{GC}},
 \label{eta-def-0}
\eea
the boundary condition Eq.~(\ref{BC-sigma}) can now be expressed into the following concise form:
\be
\frac{2}{\sinh z_0} = \eta.  \label{BC-eta}
\ee

\begin{figure}
\begin{center}
\includegraphics[width=8cm]{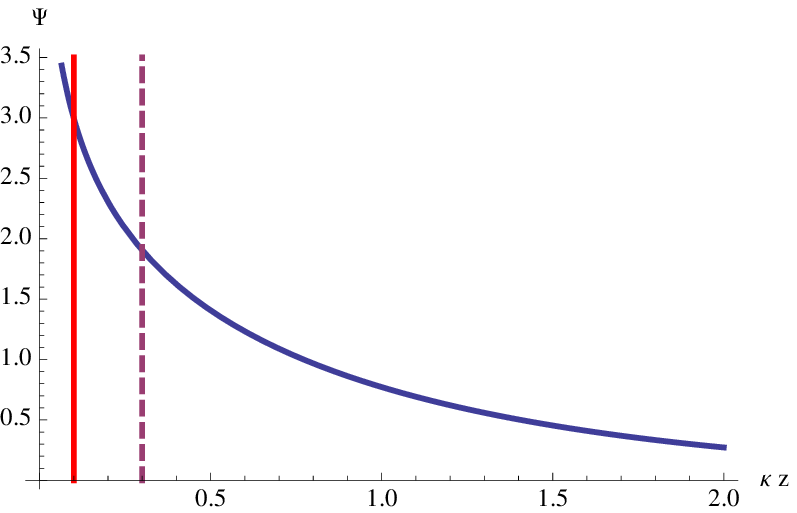}
\\
\vspace{6mm}
\includegraphics[width=8cm]{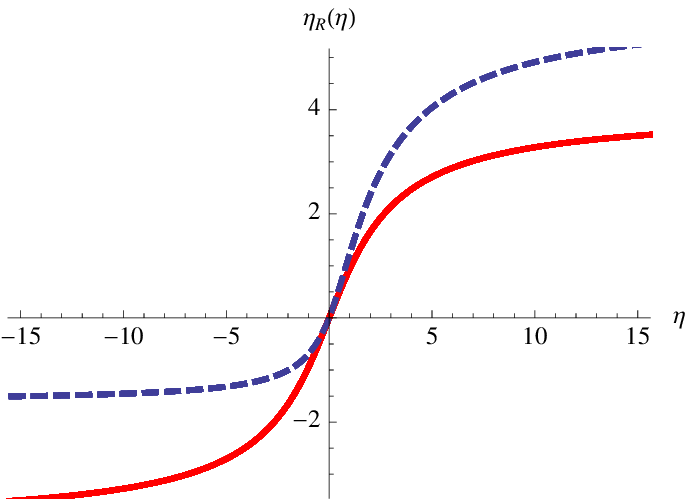}
\caption{ Up: The potential Eq.~(\ref{potential-one-plate}) can be produced by two plates with different surface charge densities and different locations.  Both satisfy the Neumann boundary condition Eq.~(\ref{BC-sigma}). The extremal possibility is an infinitely charged plate at $z_0 = 0$.  
Down: The renormalized surface charge density $\eta_R$  as a function of the bare surface charge density $\eta$ on a single uniformly charged plate.  Red solid line: a $q:-q$ symmetric electrolyte as given by Eq.~(\ref{eta_R}).  $\eta_R$ saturates at $4$ in both directions.  Blue dashed line:  a $2q:-q$ asymmetric electrolyte as given by Eq.~(\ref{relation-eta-eta_R}). $\eta_R$ saturates at $6$ and $6(2-\sqrt{3})$ in the positive and negative sides respectively.  } 
\label{potential-oneplate}
\label{effectiveQ-oneplate}
\end{center}
\end{figure}

The solution consisting of Eqs.~(\ref{potential-one-plate},\ref{BC-eta}), though simple enough, shows an interesting property.  For an arbitrary given surface charge density $\eta>0$, Eq.~(\ref{BC-eta}) can always be solved for $z_0>0$.  That is, {\em there is a one-parameter family of systems that gives the same potential Eq.~(\ref{potential-one-plate})}.  Furthermore, we can let the surface charge density $\eta$ approach infinity, then  Eq.~(\ref{BC-eta}) dictates $z_0 \rightarrow 0$.   This means that {\em Eq.~(\ref{potential-one-plate}) is the electrostatic potential produced by a plate with infinite surface charge density at the origin $z_0 = 0$}. In Fig.~\ref{potential-oneplate}, we illustrate two plates (with surface charge density determined by Eq.~(\ref{BC-eta}) ) that produce the same potential Eq.~(\ref{potential-one-plate}).  

In the near-field regime $z \ll 1$, the potential Eq.~(\ref{potential-one-plate}) reduces to the famous Gouy-Chapman solution:
\be
\Psi(z) = - 2 \log z + O(1), \quad \mbox{near field}. 
\label{Guo-Chapman-Theory}
\ee
In the far-field regime $z \gg 1$, Eq.~(\ref{potential-one-plate}) can be expanded into Taylor series of $e^{-z}$:
\be
\Psi(z) =4\, e^{- z} + O(e^{- 2 z })
= 4 \, e^{-z_0} \, e^{-\delta z} + O(e^{- 2 z }),
\label{phi-oneplate-asym}
\ee
where $\delta z = z - z_0$ is the distance to the charged plate at $z_0$. It is understood that the parameter $z_0$ is a function of $\eta$  as determined by 
Eq.~(\ref{BC-eta}).  On the other hand, in the linear PB theory, a plate with surface charge density $\eta_R$ at the same location $z_0$ produces at $z$ a potential 
\be
\Psi^{\rm linear} (z) = \eta_R \, e^{-\delta z}.  
\label{solution-linear-BP}
\ee
Following the proposal by Alexander {\it et. al.} \cite{charge-RG-Alexander-JCP-1984}, we identify the far-field asymptotics Eq.~(\ref{phi-oneplate-asym}) of the nonlinear theory with the linear theory Eq.~(\ref{solution-linear-BP}), and define the renormalized (or effective) charge density $\eta_R$.   Using Eq.~(\ref{BC-eta}) to eliminate $z_0$ in favor of $\eta$, we find the renormalized surface charge density $\eta_R (\eta)$ as a function of the bare surface charge density $\eta$: 
\be
\eta_R(\eta) =\frac{2 \eta}
{ 1+ \sqrt{1+(\eta/2 )^2} },  
\label{eta_R}
\ee
which is illustrated by the red solid curve in the right panel of Fig.~\ref{effectiveQ-oneplate}.  In the weakly charged limit $\eta \ll 1$, the renormalized charge density $\eta_R$ is just the bare one $\eta_R \rightarrow \eta $; in the strongly charged limit $\eta \gg 1$, $\eta_R$ saturates at $\eta_R( + \infty) = 4$.  
$\eta_R$ saturates at $- 4$ when $\eta$ is negative and large.  More generally, Eq.~(\ref{eta_R}) is invariant under the charge-inversion transformation $(\eta, \eta_R) \rightarrow (- \eta, - \eta_R)$, in accordance with the fact that the electrolyte is {\em symmetric.}   The practice of using linear theory with a renormalized, i.e. effective, surface charge density in the far field is usually called {\em charge renormalization} following the seminal work by Alexander {\it et al}.  Here the most striking property is that $\eta_R$ saturates at a finite value in the strongly charged limit.  {\em It shows that electrolytes are able to screen infinitely charged objects within finite distance.}  This should be regarded as one fundamental property of the Poisson-Boltzmann theory.  

In general, the renormalized surface charge density depends on the shape of the charged object.  It also depends on the properties of electrolyte.   For example, our analyses in this work show that inside a $2q:-q$ asymmetric electrolyte, a positively charged plate produces a dimensionless potential 
\be
\Psi_{2q:-q}^+(z) = \log \frac{1+ 4 \, e^{-z} + e^{-2z}}{(1-e^{-z})^2},
\label{potential-one-plate-1}
\ee
which diverges logarithmically at $z = 0$.  A negatively charged plate in a $2q:-q$ electrolyte, on the other hand, produces a potential 
\be
\Psi_{2q:-q}^-(z) = \log \frac{1- 4 \, e^{-z} + e^{-2z}}{(1+ e^{-z})^2}, 
\label{potential-one-plate-2}
\ee
which diverges at $z_m = \log (2 + \sqrt{3})$.  These potentials were discovered by Andrietti {\it et al} in 1976, but has remained largely unknown since then.  The far-field asymptotics of these potentials are given by 
\be
\Psi_{2q:-q}^{\pm} (z) \sim  \pm 6\, e^{-z} + O(e^{-2z}). 
\label{potential-one-plate-asym-1}
\ee
Carrying out a similar analysis as in the case of symmetric electrolyte, we find that the relation between the renormalized surface charge density $\eta_R$ and the bare surface charge density $\eta$ as
\be
\frac{36 \eta _R \left(\eta
   _R+6\right)}{\left(6-\eta _R\right)
   \left(\eta _R^2 + 24 \eta_R +36\right)} 
   = \eta, 
 \label{relation-eta-eta_R}
\ee
which is illustrated by the dashed blue curve in Fig.~\ref{effectiveQ-oneplate}.
It is interesting to note that for $\eta >0$, $\eta_R$ saturates at $\eta_R(\infty) = 6$, while for $\eta <0$, $\eta_R$ saturates at $\eta_R(-\infty) =  -  6(2 - \sqrt{3} )\approx - 1.6077$.  The fact $\eta_R( \eta) \neq - \eta_R( - \eta)$ shows that the electrolyte is indeed {\em asymmetric}.  There does not seem to exist simple analytic result for the one plate problem in more general $mq:-nq$ asymmetric electrolytes.

The finiteness of electrostatic potential away from an infinitely charged object in electrolyte is a short scale property of the PB equations, and therefore is expected to hold independent of the large scale geometry of the charged objects.  Indeed the phenomenon of charge renormalization was first discovered in numerical studies of spherical colloids \cite{charge-RG-Alexander-JCP-1984}.  It was found there that  the effective interaction between charged objects appears to remain finite when the surface charge density is scaled up.  However that has not been any analytic result showing that this is indeed true \footnote{In the real world, of course, there is no infinitely charged object.  Furthermore, all ions have some effective hard core radii, within which other ions can not penetrate.  Hence there is an upper bound of bulk charge density that can be achieved in any given electrolyte, which is of order of $q/a^3$, where $a$ is the ion diameter.   Short scale details ignored by PB theory must come into play, therefore, when the surface charge density becomes sufficiently high.  We shall however not worry about this issue in the present work.  
The saturation of the renormalized surface charge density happens around $\eta \sim 1$.  The physical surface charge density scales as 
$\sigma^* \sim \eta \kappa /\beta q$, see Eqs.~(\ref{eta-def-0}).   For a typical value of Debye length $\ell_{DB} = \kappa^{-1} = 10^{-8} m$, with monovalent ions, we have $\sigma^* \approx 0.01 e/nm^2$ (with $e = 1.6\times 10^{-19}$ is the charge of one electron), which is two orders of magnitude smaller than the surface charge density of e. g. DNA.  In the experiment by Tata {\it et al}, the Debye length is about $500 nm$, which corresponds to a cross-over charge density $\sigma^* \approx 2 \times 10^{-4} e/nm^2$, which is three orders of magnitude smaller than the colloidal surface charge density in the same experiment.  Therefore, there is a large window of length scales where the phenomenon of charge renormalization can be observed.}.  For the case of spherical colloids, even the potential by a single colloid is unknown.  In general, lack of analytic results is one major obstacle to understanding of the physics of charged colloids. 


\begin{figure}
\begin{center}
\includegraphics[width=8.5cm]{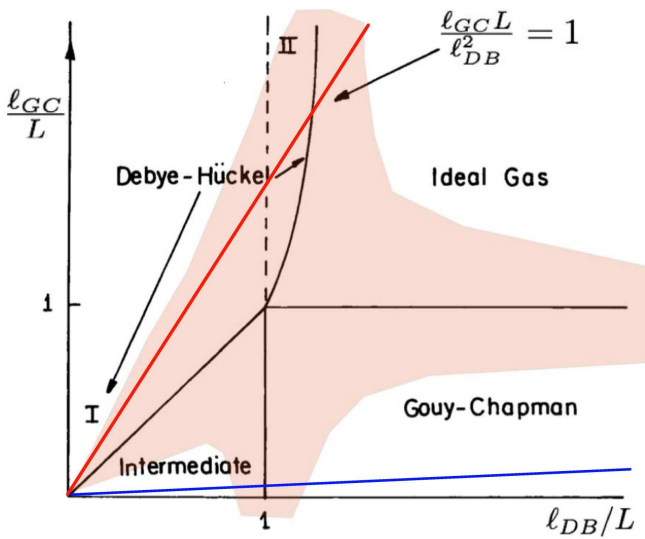}
\caption{The two dimensional parameter space for the two plate problem.   There are four regimes where simple asymptotic results can be obtained without solving the nonlinear PB.  Our solution in terms of Weierstrass elliptic functions reduces to the known asymptotic results in the corresponding regimes, and works in the cross-over regions shown shaded in the figure as well. Part of this plot was taken from the review by Andelman \cite{Andelman-PBequation}.  The blue line: two strongly charged plates with a variable distance.  The red line: two weakly charged plates with a variable distance. }
\label{four-regimes}
\end{center}
\vspace{-5mm}
\end{figure}

The problem of two charged plates is the simplest toy model for effective interactions between charged colloids.  For two likely charged inside a symmetric electrolyte, a formal solution of the nonlinear PB was obtained in terms of elliptic integrals in the classical monograph by Verwey and Overbeek \cite{Verwey-Overbeek} as early as 1940's.  From this, asymptotic results can be obtained in four regimes shown in Fig.~\ref{four-regimes}.  For a detailed discussion, see the review by Andelman \cite{Andelman-PBequation}.  There however has been no analytic result in the cross-over regions, shaded in Fig.~\ref{four-regimes}.   In the present work, we shall analytically calculate the electrostatic potential profile and the interaction between the two identically charged plates, both in symmetric and in $2:1$ asymmetric electrolytes.  By expressing the electrostatic potential in terms of Weierstrass elliptic functions, we are able to calculate the interaction between two plates in the {\em whole parameter space}.  We derive exact results in various asymptotic regimes as well as cross-over regimes.  In particular, we find analytic expressions for the interactions between two infinitely charged plates, both in symmetric and asymmetric electrolytes.  These results constitute the most direct demonstration of charge renormalization.  We also find that the potential by a single plate in a $2q:-q$ asymmetric electrolyte in terms of elementary functions, as discussed above.  Last but not least, our result also demonstrate how bi-valence counter-ions dramatically reduces the interaction between two likely charged plates.  It is important to note that our solution should be equivalent to one by Verwey and Overbeek \cite{Verwey-Overbeek}.  In fact, it is known that all elliptic functions can be expressed in terms of Weierstrass elliptic functions.  The representation in terms of Weierstrass elliptic functions is however particularly convenient because of their well known analytic structure.  It allows us to deduce many analytical properties of the PB equation for two-plates geometry.   

In recent years, there has been much discussion on the physics of charged colloids beyond Poisson-Boltzmann theory \cite{RevModPhys.74.329,Levin-charge-review,Boroudjerdi2005129,Messina-review-electrostatics}.  Of particular interests are the counter-intuitive possibilities such as charge inversion, like-charge attraction, etc.  Partly due to the lack of reliable analytic results, however, the situation remains murky after more than a decade of debates.  In this work, we shall limit ourself inside the domain of Poisson-Boltzmann theory.  It is our belief that the physics {\em beyond}  PB can be understood only after we understand the physics {\em within} PB theory.   

The remaining of this paper is organized as follows.  In Sec.~\ref{sec:elliptic_function} we construct the solution to PB in terms of Weierstrass elliptic function, both in symmetric electrolytes and in $2q:-q$ asymmetric electrolytes.   We also express the interaction between two plates in terms of the first integral of the PB equation.   In Sec.~\ref{sec:one-plate}, we discuss the one plate problems,  analytically calculate the electrostatic potential and the renormalized surface charge density as a function of the bare surface charge density.  In Sec.~
\ref{sec:two-plates-infinite}, we study the problem of two infinitely charged plates.  We obtain the exact asymptotics both in the large separation regime and in the small separation regime, and find that multivalent counter-ions significantly reduce the repulsion between two likely charged plates.  We also obtain analytic results for arbitrary separation.   In Sec.~\ref{sec:two-plates-weak} we study two weakly charged plates and obtain exact asymptotics both in the large separation limit and in the small separation limit.  In Sec.~\ref{sec:general-case} we analyze the case of two strongly charged plates and briefly discuss the most general case where the surface charge density is neither strong nor weak.  Finally in Sec.~\ref{sec:conclusion} we summarize our result and discuss possible future directions. 
 In Sec. \ref{app:derivation-PB}  we present two different derivations of Poisson-Boltzmann equation in symmetric electrolytes, and discuss some subtleties in each derivation.  In Sec. \ref{app:general_results} we discuss some general properties of PB equation with the geometry of two parallel charged plates.  The pressure between plates naturally emerges as the first integral of PB equation. In Sec. \ref{app:asymmetric} we derive PB for the case of asymmetric electrolytes and discuss some of its general properties.   

\section{Solving Poisson-Boltzmann Equation}
\label{sec:elliptic_function}
Inside a symmetric electrolyte, the Poisson-Boltzmann equation for the dimensionless electrostatic potential $\Psi = \beta q \phi$ between two identical charged plates is given by 
\be
-  \Psi''(z) + \sinh \Psi(z) = 0.  
\label{saddle-eqn-1}
\ee
Let $\epsilon, \epsilon'$ be the dielectric constants for the electrolyte and for the plates.  In the limit $\epsilon \gg \epsilon'$, the relevant electrostatic boundary condition reduces to that of Neumann: 
\be
\left. \frac{\partial \Psi}{\partial n} \right|_{\rm plates} = \eta,
\label{Neumann-BC}
\ee
where the normal director points into the plates.  It is easy to obtain the following first integral:
\be
{\alpha} = \cosh \Psi (z)- \frac{1}{2} (\Psi'(z))^2
= \cosh \Psi(MP), 
\label{first-integral}
\ee
where ``MP'' stands for ``middle point''.  The net interaction between two plates only depends on the constant $\alpha$ and is given by 
\be
P_{\rm net} = 2 n\, T (\alpha - 1) 
= \frac{T\,\delta \alpha}{ 4 \pi \ell_{DB}^2 \lambda_{Bj}}, 
\label{P_net}
\ee
where $\delta \alpha = \alpha - 1$, and  $2 \,n$ is the total density of ions in the bulk. $\alpha > 1$ for two likely charged plates.  $\alpha \rightarrow 1$ as the distance between plates becomes large.  For detailed derivations of all these results, see Sec. \ref{app:derivation-PB}. 

\begin{figure}
\begin{center}
\includegraphics[width=6cm]{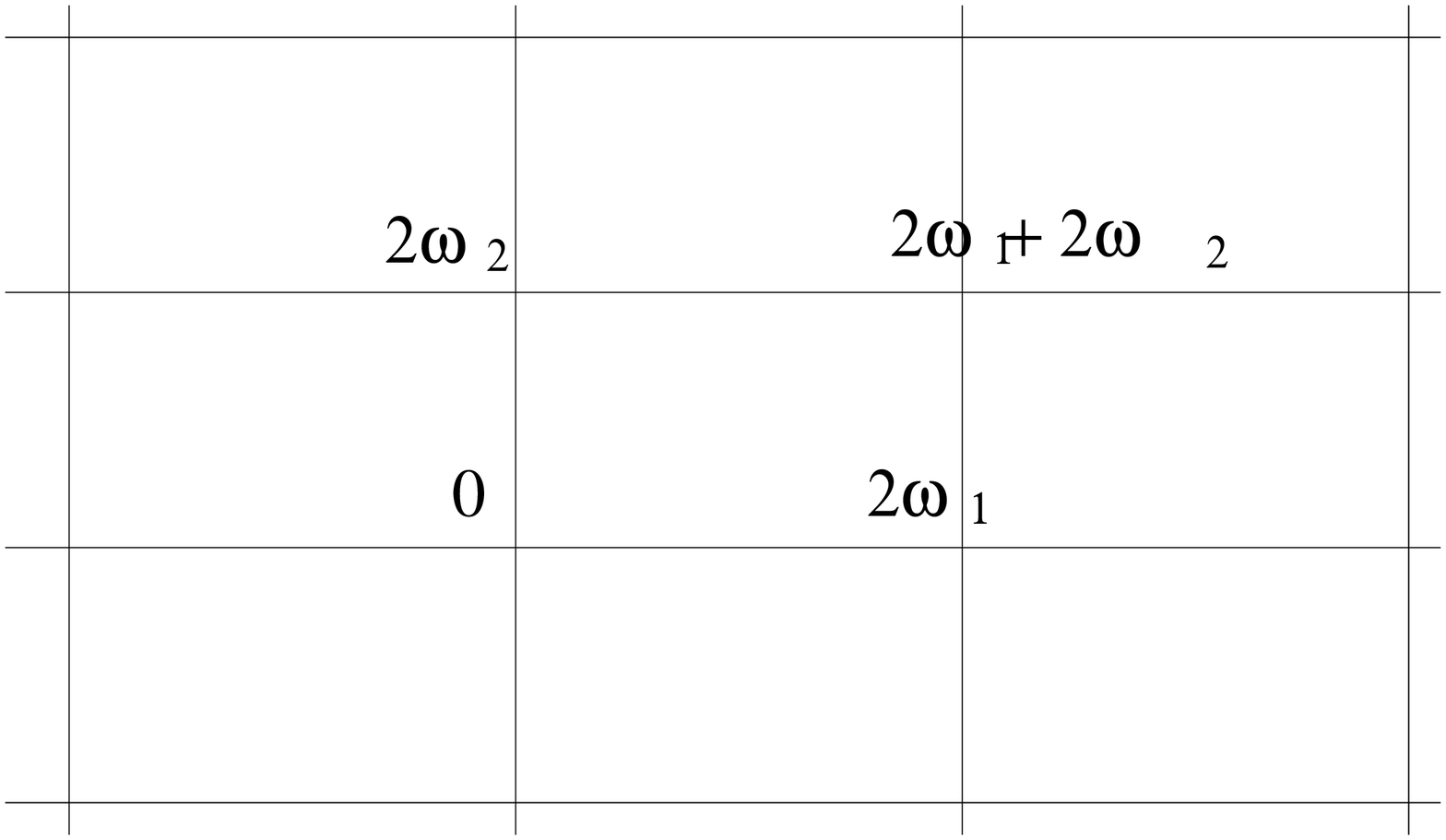}
\\
\vspace{5mm}
\includegraphics[width=7cm]{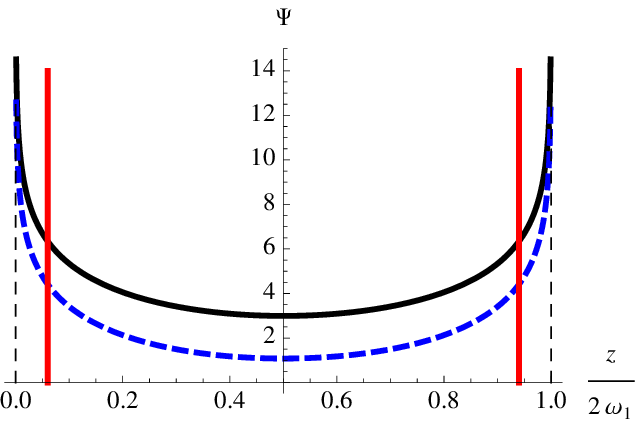}
\caption{(Up) The periodic lattice structure of the singularities of $\wp(z,g_2,g_3)$ in the complex plane.  Each grid point is a pole of second order.  For $\alpha >1$, $2 \omega_1$ is real, while $2 \omega_2$ is purely imaginary.  
(Down) Black solid curve: Eq.~(\ref{Psi-p}) as a function of normalized real coordinate $z/2 \omega_1$ with $\alpha > 1$. $z = 0, 2 \omega_1$ are logarithmic singularities; $z = \omega_1$ is mid point between plates.  Here $\alpha = 10$, corresponding to a period $2 \omega_1 \approx 1.40761 \, \ell_{DB}$.  Two dashed lines are two maximally separated plates with infinite surface charge density, for given pressure $\alpha$.  Two solid (red) lines are two plates with finite surface charge density that gives the same pressure.   The parameter $z_0$ in Eq.~(\ref{z0A-1}) is the distance between the dashed line and the solid line, i.e. it is the distance between the plate and the logarithmic singularity nearby.  Blue dashed curve: the potential Eq.~(\ref{Psi-asym}) inside $2q:-q$ asymmetric electrolytes with $\alpha = 1$, and $2 \omega_1 = 4.52566 \,\ell_{DB}$.  }
\label{wp-1}
\label{potential-plot}
\end{center}
\end{figure}

Now introducing a new function $\wp(z)$ through 
\be
4 \wp(z) + 2 \alpha / 3 = e^{\Psi(z)}, 
\label{y-Psi}
\ee
the first integral Eq.~(\ref{first-integral}) is transformed into 
\be
(\wp')^2 = 4\, \wp^3 - g_2 \, \wp - g_3,
\label{Elliptic-ODE}
\ee
where 
\bea
g_2 =   \frac{\alpha ^2}{3} - \frac{1}{4} , \quad
g_3 =  \frac{\alpha ^3}{27} - \frac{\alpha}{24}. 
\label{g-A}
\eea
Eq.~(\ref{Elliptic-ODE}) is the differential equation satisfied by the doubly periodic {\em Weierstrass elliptic function} $\wp(z;g_2,g_3)$, with $g_2$ and $g_3$ two of its {\em invariants}.   Below we invoke some of its well known properties without any proof.  The readers are referred to the classic monograph by Whittaker and Waston \cite{Whittaker-Watson} for relevant details.  The function $\wp(z;g_2,g_3)$ can be explicitly represented as a double series in $z$:
\bea
\wp(z) = \frac{1}{z^2} + {\sum_{m,n}} '
&&\left[ 
\frac{1}{ (z - 2 m \omega_1 - 2 n \omega_2 )^2 } 
\right. \nonumber\\
&-& \left. \frac{1}{( 2 m \omega_1 + 2 n \omega_2)^2} 
\right],
\label{P-series}
\eea
where the prime in the summation means exclusion of the term with $m = n = 0$.  $\wp(z)$ is a meromorphic function of $z$, generally treated as a {\em complex variable}.  $2 \omega_1$ and $2 \omega_2$ (complex numbers in general) are the two periods and are related to the two invariants via 
\begin{subequations}
\label{g2g3-def}
\bea
g_2( \omega_1,\omega_2) 
&=& 60  {\sum_{m,n}} ' \frac{1}{( 2 m \omega_1 + 2 n \omega_2)^4},\\
g_3(\omega_1,\omega_2)
&=& 140  {\sum_{m,n}} ' \frac{1}{( 2 m \omega_1 + 2 n \omega_2)^6}.  
\eea
\end{subequations}
All series involved here are absolutely convergent.  The function $\wp(z;g_2,g_3)$ is completely determined by two invariants $g_2,g_3$, or equivalently, its two periods $2\omega_1, 2\omega_2$.  In our case, they depend on only one parameter $\alpha$, see Eq.~(\ref{g-A}).   One can easily check using Eq.~(\ref{P-series}) that $\wp(z)$ is indeed periodic with respect to two periods $2 \omega_1$ and $2 \omega_2$:
\be
\wp(z) = \wp(z+ 2 \omega_1) = \wp(z + 2 \omega_2). 
\ee
Furthermore, two periods are guaranteed to be linearly independent as two dimensional vectors in the complex plane \cite{Whittaker-Watson}.  From Eq.~(\ref{P-series}), $z = 0$ is a second order pole of $\wp(z)$.  By periodicity, $\wp(z)$ has an infinite number of second order poles $z_{m,n} = 2 m \omega_1 + 2 n \omega_2$, which form a 2D oblique lattice.  These are the only singularities of $\wp(z)$.  In the remaining of this work, we shall focus on the case of two likely charged plates where $\alpha >1$.  For this case, it can be explicitly shown that the lattice of singularities is rectangle.  Hence $ 2\omega_1$ and $2 \omega_2$  can be chosen to be real and purely imaginary respectively.  This lattice is illustrated in Fig.~\ref{wp-1}.  The case with $\alpha <1$ (two oppositely charged plates) is slightly more complicated and will not be discussed here.  For $\alpha>1$, the real half period $\omega_1$ is related to $\alpha$ through the two invariants Eq.~(\ref{g-A}): 
\bea
\omega_1 (\alpha) &=& \Upsilon_{\rm WHP}
\left( g_2(\alpha),g_3(\alpha)\right)
\nonumber\\
&=&  \Upsilon_{\rm WHP}
 \left( 
\frac{\alpha ^2}{3} - \frac{1}{4},
\frac{\alpha ^3}{27} - \frac{\alpha}{24}
\right), 
\label{omega-1-alpha}
\eea
where the subscript ``WHP'' stands for ``Weierstrass Half period''.  This function can be conveniently computed using Wolfram Mathematica 7.  \footnote{The relevant Mathematica function is {\it WeierstrassHalfPeriods}$[\{g_2,g_3\}]$.  It returns two half periods for a given elliptic function.  However, since for $\alpha >1$ one of the periods is real and the other is purely imaginary, it is straightforward to extract the real one which is what we want here. }

From now on we shall consider $z$ as a real variable, since this is ultimately what we  concern about.  Inverting Eq.~(\ref{y-Psi}), we find that the dimensionless potential $\Psi$ is given by
\be
\Psi(z) = \log \left[ 
4 \,\wp \left(z; \frac{\alpha ^2}{3} - \frac{1}{4},
\frac{\alpha ^3}{27} - \frac{\alpha}{24}
\right) + \frac{2 \alpha} {3} 
\right],    \label{Psi-p}
\ee
which has logarithmic singularities at $z = 0, \pm 2\omega_1, \pm 4 \omega_1, \ldots$.  These singularities have the same nature as the one in the solution for the one-plate problem.  In fact, Eq.~(\ref{potential-one-plate}) is an elliptic function with $(2 \omega_1,2 \omega_2) = (+\infty, 2 \pi i)$.  In Fig.~\ref{potential-plot}, we plot this potential in one period along the real axis $z \in \left(0, 2 \omega_1 \right)$ for a particular value of $\alpha = 10$.  It is positive everywhere, and diverges logarithmically at $z = 0, 2 \omega_1 \approx  1.4076$.  It therefore describes the potential between two {\em positively charged plates}.  
The potential between two negatively charged plates is simply the negative of Eq.~(\ref{Psi-p}), since the saddle point equation Eq.~(\ref{saddle-eqn-1}) is invariant under transformation $\Psi  \rightarrow - \Psi$.  This symmetry no longer holds in an asymmetric electrolyte, as we will see shortly after.  One possible loci for two positively charged plates are illustrated by  the solid lines in Eq.~(\ref{potential-plot}).  Like the case of a single charged plate, we can continuously change the loci of the plates as well as the surface charge densities in such a way that the Neumann BC is always satisfied and the constant $\alpha$ is fixed.  That is, there is a one parameter family of problems that the yield the identical potential profile.  Since the potential between two charged plates must be a smooth function, the two singularities $0, 2 \omega_1$, must be {\em outside} two plates.  This means that for a given pressure $\alpha$, there is a {\em maximum separation} between two plates, which is precisely the real period $2 \omega_1$.  {\em The surface charge density yielding this interaction at this maximal separation is infinity}, see the dashed vertical lines in Fig.~\ref{potential-plot}. 

Similar to the one plate problem, we let the plate on the left to be at $z_0$.  Since the other plate carries the same charge density, and since $\wp(z)$ is an even function, the other plate must be located at $z = 2 \omega_1 - z_0$.   The values of $\alpha , z_0$ should then be determined by two boundary conditions Eq.~(\ref{Neumann-BC}).  

By our choice of coordinate system, $z = \omega_1$ is the mid point between two plates.  It is known that $\wp$ at this point satisfies the following equation \cite{Whittaker-Watson}:
\be
4 t^3 - g_2 t -g_3  = 0,  
\ee
with $g_2,g_3 $ given by Eq.~(\ref{g-A}).  This equation admits three real roots if $\alpha >1$:
\be
- \frac{\alpha}{6}, \quad
 \frac{1}{12} 
\left(-3 \sqrt{\alpha ^2-1} + \alpha\right), \quad
 \frac{1}{12} 
\left(3 \sqrt{\alpha ^2-1} + \alpha\right). 
\label{wp-omega1}
\ee
Only the last one gives a positive potential $\Psi$ through Eq.~(\ref{Psi-p}).  It is also easy to check that this solution satisfies Eq.~(\ref{first-integral}).  

We note in passing that from the electrostatic potential Eq.~(\ref{Psi-p}), one can straightforwardly calculate the density of counter-ions and co-ions between the plates.  We shall however not elaborate on this issue in the present work.  

\subsection{$2q:-q$ Asymmetric Electrolyte}
\label{sec:asymmetric}
Let positive ions and negative ions carry charge of $2 q$ and $-q$ respectively.  To keep overcall charge neutrality, their bulk ion densities are $n /2$ and $n$ respectively.  The Debye length for a $2q:-q$ electrolyte is \cite{Stat-Mech:Landau} 
\be
\ell_{DB} = \sqrt{\frac{\epsilon}{3 n \, \beta q^2}}.   
\label{Debye-asym}
\ee
The Poisson-Boltzmann equation in a $2q:-q$ electrolyte is
\be
-  \partial_z^2 \Psi + \frac{1}{3} e^{\Psi} - \frac{1}{3} e^{-2 \Psi} = 0.  
\label{saddle-asym}
\ee
Linearization of this equation leads to Eq.~(\ref{PB-linear}), as it should be. \footnote{Note that we have rescaled the coordinate by the Debye length. } The first integral of Eq.~(\ref{saddle-asym}) is:
\bea
\alpha &=& - \frac{1}{2} (\partial_z \Psi)^2 
+ \frac{1}{3}e^{\Psi} + \frac{1}{6} e^{-2 \Psi}
\nonumber\\
&=& \frac{1}{3}e^{\Psi(MP)} + \frac{1}{6} e^{-2 \Psi(MP)}.  
\label{A-asymmetric}
\eea
For likely charged plates, $\alpha \geq 1/2$.  $\alpha$ converges to $1/2$ as the distance between plates becomes large. The net interaction between two plates is given by 
\be
P_{\rm net} 
= 3 nT\left( \alpha - \frac{1}{2} \right) 
= \frac{T \delta \alpha}{4 \pi \ell_{DB}^2 \lambda_{Bj}},
\label{P-alpha-asym}
\ee
where $\delta \alpha = \alpha - 1/2$.  For detailed derivation of the above results, see Sec. \ref{app:asymmetric}.   


Introducing a variable $ \wp(z)$ via
\be
6  \, \wp(z) + \alpha = \exp \Psi,
\label{w-psi-asym}
\ee 
we find that Eq.~(\ref{A-asymmetric}) reduces to Eq.~(\ref{Elliptic-ODE})
with two invariants given by 
\bea
g_2 =   \frac{\alpha ^2}{3} ,\quad
g_3 =  \frac{\alpha ^3}{27} - \frac{1}{108}. 
\label{g-A-asym}
\eea
Inverting Eq.~(\ref{w-psi-asym}), we find that the electrostatic potential is given by
\be
\Psi = \log \left[ 
6 \, \wp\left(z; \frac{\alpha ^2}{3},
 \frac{\alpha ^3}{27} - \frac{1}{108} \right) + \alpha \right]. 
\label{Psi-asym}
\ee
For real $z$ and $\alpha > 1/2$, this potential is always positive.  In Fig.~\ref{potential-plot} we plot Eq.~(\ref{Psi-asym}) within one period on the real axis for the particular case of $\alpha =1 $.  {\em Therefore Eq.~(\ref{Psi-asym}) with $z$ on the real axis describes the electrostatic potential between two positively charged plates, as long as $\alpha > 1/2$}.  The real period $2 \omega_1$ as a function of $\alpha$ is given by
\be
2 \omega_1(\alpha) = 
2\,\Upsilon_{\rm WHP} \left( 
\frac{\alpha ^2}{3} ,
\frac{\alpha ^3}{27} - \frac{1}{108} 
\right).  
\label{omega-1-alpha-asym}
\ee

Because the electrolyte is asymmetric, the saddle point equation Eq.~(\ref{saddle-asym}) is not invariant under the transformation $\Psi \rightarrow - \Psi$.   The potential by negatively charged plates has to be studied separately.  We shall come back to this problem below.  

\section{One plate problem}
\label{sec:one-plate}
For the case of symmetric electrolyte, $\alpha \rightarrow 1$ as the distance between plates approaches infinity, see Eqs.~(\ref{first-integral}, \ref{P_net}).   The two invariants Eq.~(\ref{g-A}) reduce to 
\be
g_2 \rightarrow \frac{1}{12},\quad
g_2 \rightarrow - \frac{1}{216}. 
\label{g2-g3-limit}
\ee
Using Wolfram Mathematica 7, we can explicitly check 
\be
2 \omega_1 \rightarrow + \infty, \quad
2 \omega_2 \rightarrow 2 \pi i .
\ee
The electrostatic potential Eq.~(\ref{Psi-p}) then becomes 
\bea
\Psi_{q:-q}(z) &=& 
\log \left[ 
4 \wp\left(z;\frac{1}{12},-\frac{1}{216}\right)
+\frac{2}{3}
 \right]. 
 \label{potential-one-plate-2-1}
\eea
At a given finite $z$, this potential is completely due to the plate near $z = 0$, since the other plate is infinitely far away.  On the other hand, we already know that the potential by one charged plate inside a symmetric electrolyte is given by Eq.~(\ref{potential-one-plate}).  Comparing it with Eq.~(\ref{potential-one-plate-2-1}) we find the following useful identity: 
\be
\wp\left(z;\frac{1}{12},-\frac{1}{216}\right)
= \frac{1}{4} \left( \frac{1+ e^{-z}}{1-e^{-z}}\right)^2 - \frac{1}{6}.  
\label{wp-limit}
\ee  

\subsection{$+$ plate in $2q:-q$ electrolyte}
Now consider a $2q:-q$ asymmetric electrolyte.  When two plates are widely separated, $\alpha \rightarrow 1/2$, see Eqs.~(\ref{A-asymmetric}, \ref{P-alpha-asym}).  Interestingly, the two invariants Eqs.~(\ref{g-A-asym}) approach the same limit Eq.~(\ref{g2-g3-limit}).  Using Eq.~(\ref{wp-limit}) and Eq.~(\ref{Psi-asym}), we find that inside a $2q :-q$ asymmetric electrolyte, the electrostatic potential of a positively charged plate is given by 
\bea
\Psi_{2q:-q}^+(z) &=&  
\log \left[ 
6 \,\wp\left(z;\frac{1}{12},-\frac{1}{216}\right)
+ \frac{1}{2}
 \right] 
 \nonumber\\
&=& 
\log \frac{1+ 4 \, e^{-z} + e^{-2z}}{(1-e^{-z})^2},  
\label{potential-one-plate-1-1}
\eea
which we have already presented in Eq.~(\ref{potential-one-plate-1}).   It is easy to check explicitly that Eq.~(\ref{potential-one-plate-1-1}) indeed satisfies PB Eq.~(\ref{saddle-asym}).  Furthermore for $z>0$, Eq.~(\ref{potential-one-plate-1-1}) is positive everywhere.   

Let the plate be at $z_0$ with a surface charge density $\eta>0$, the boundary condition Eq.~(\ref{Neumann-BC}) dictates
\be
- \frac{d  }{dz_0}\Psi_{2q:-q}^+(z_0) = 
\frac{3
   \left(e^{z_0}+1\right)}{\left(e^{z_0}-1\right)
   (\cosh z_0 + 2)}
   = \eta.  
   \label{z_0-eta-asym-1}
\ee 
On the other hand, the far field asymptotics ($z \gg 1$) of Eq.~(\ref{potential-one-plate-1-1}) is
\be
\Psi_{2q:-q}^+ (z)\approx 6 \,e^{-z}
= 6 \, e^{-z_0} \, e^{- \delta z}, 
\quad   \delta z = z - z_0.
\label{Psi-limit-asym-1}
\ee
Comparing this with the lienarized PB theory Eq.~(\ref{solution-linear-BP}), we find that the renormalized surface charge density $\eta_R$ is given by
\be
\eta_R = 6 \, e^{-z_0} > 0 , 
\label{eta_R-z_0-asym}
\ee
where $z_0$ is in turn related to $\eta$ by Eq.~(\ref{z_0-eta-asym-1}).  In the limit $\eta \rightarrow \infty$, $z_0 \rightarrow 0^+$ according to Eq.~(\ref{z_0-eta-asym-1}), and $\eta_R$ saturates at $6$.  Therefore Eq.~(\ref{potential-one-plate-1-1}) {\em gives the potential by a positive infinitely charged plate located at } $z_0 = 0$.  For finite value of $\eta$, we can use Eq.~(\ref{eta_R-z_0-asym}) to eliminate $z_0$ in favor of $\eta$ in Eq.~(\ref{z_0-eta-asym-1}), and obtain the relation between $\eta_R$ and $\eta$:
\be
\frac{36 \, \eta _R \left(\eta
   _R+6\right)}{\left(6-\eta _R\right)
   \left(\eta _R^2 + 24 \, \eta_R +36\right)} 
   = \eta,
 \label{relation-eta-eta_R-1}
\ee
which we already presented in Eq.~(\ref{relation-eta-eta_R}) in Sec.~\ref{sec:intro}.  It is easy to verify using  this relation that $\eta_R \rightarrow 6$ as $\eta \rightarrow \infty$.  

\subsection{$-$ plate in $2q:-q$ electrolyte} 
Rather than presenting the derivation that leads to the correct solution, we directly show the result here.  The function 
\bea
\Psi^{-}_{2q:-q} (z) &=& 
\Psi^{+}_{2q:-q} (z +  i \pi) 
\nonumber\\
&=& 
\log \frac{1- 4 \, e^{-z} + e^{-2z}}{(1 + e^{-z})^2}, 
\label{potential-one-plate-2-2}
\eea
also satisfies the differential equation Eq.~(\ref{saddle-asym}), and is negative as long as $z > \log (2 + \sqrt{3}) \approx 1.317$.  It has a logarithmic singularity at $z = \log (2 + \sqrt{3})$.  It therefore describes the potential by a negatively charged plate inside a $2q:-q$ electrolyte.  

Let the plate be at $z_0$ with a surface charge density $\eta < 0$, the boundary condition Eq.~(\ref{Neumann-BC}) dictates
\be
- \frac{d  }{dz_0}\Psi_{2q:-q}^- (z_0) = 
- \frac{3 \left(e^{z_0} - 1\right)}{\left(e^{z_0}+1\right)
   (\cosh z_0 - 2)}
   = \eta.  
   \label{z_0-eta-asym-2}
\ee 
In the limit $\eta \rightarrow - \infty$, we have $z_0 \rightarrow \cosh^{-1} 2 = \log (2 + \sqrt{3})$.  Therefore Eq.~(\ref{potential-one-plate-2-2}) describes the potential by a plate with $\eta = - \infty$ at $z_0 = \log (2 + \sqrt{3})$.  In the far field $z \gg 1$, the asymptotic behavior of the potential Eq.~(\ref{potential-one-plate-2-2}) is
\be
\Psi_{2q:-q}^- (z)\approx 6 \,e^{-z}
= - 6 \, e^{-z_0} \, e^{- \delta z}, 
\quad   \delta z = z - z_0.
\label{Psi-limit-asym-2}
\ee
Comparing this with the lienarized PB theory Eq.~(\ref{solution-linear-BP}), we find that the renormalized surface charge density $\eta_R$ is given by
\be
\eta_R = - 6 \, e^{- z_0} <0, 
\label{eta-z_0-2}
\ee
where $z_0$ is in turn related to $\eta$ by Eq.~(\ref{z_0-eta-asym-2}).  Eliminating $z_0$ in favor of $\eta$ in Eq.~(\ref{z_0-eta-asym-2}) using Eq.~(\ref{eta-z_0-2}), we find the relation between the renormalized surface charge density $\eta_R$ and the bare density $\eta$ (both negative in this case),  which is identical to Eq.~(\ref{relation-eta-eta_R-1}).    $\eta_R$ as a function of $\eta$ is illustrated in Fig.~\ref{effectiveQ-oneplate}, for both positive and negative $\eta$.   $\eta_R$  saturates at $ -6/(2 + \sqrt{3}) \approx 1.6077$ as $\eta \rightarrow - \infty$, and saturates at $6$ as $\eta \rightarrow + \infty$. 

\section{Two Widely Separated Plates with Arbitrary Charges} 
\label{sec:large-separation}
The solution to the one plate problem help us understanding the asymptotics of the two plate problem where the separation is much longer than the Debye length, $L \gg \ell_{DB}$.  The potential in the middle is approximately the sum of that of two isolated plates.  In the middle of plates, therefore, 
\be
\Psi(MP) =  2 \eta_R(\eta) \, e^{-L/2}.  
\ee
Using Eqs.~(\ref{first-integral}, \ref{P_net}), or Eqs.~(\ref{A-asymmetric}, \ref{P-alpha-asym}), and the fact that $\Psi(MP) \ll 1$,  the net interaction between two plates is approximately 
\bea
\delta \alpha &=& \frac{1}{2} \Psi(MP)^2 = 2 \eta_R(\eta)^2 e^{-L/\ell_{DB}}, \nonumber\\ 
P_{\rm net} &=& 
\frac{T \eta_R(\eta)^2}{2 \pi \ell_{DB}^2 \lambda_{Bj}}
 e^{-L/\ell_{DB}}, 
\label{P-large-separation}
 \eea
where we have restored the physical unit for length, while $\eta_R$ is given by Eq.~(\ref{eta_R}) for symmetric electrolytes and by Eq.~(\ref{relation-eta-eta_R}) for $2q:-q$ asymmetric electrolytes.    Note that this result be obtained from the corresponding result Eq.~(\ref{P-DH}) in the Debye-H\"{u}ckel regime by replacing the bare surface charge density $\eta$ with the renormalized one $\eta_R$ in the regime where $L \gg \ell_{DB}$.   This result applies to the whole region to the left of the vertical line $\ell_{DB}/L = 1$ in Fig.~\ref{four-regimes}, which includes the intermediate regime as well as a major part of the Debye-H\"{u}ckel regime.

\section{Two Infinitely Charged Plates}
\label{sec:two-plates-infinite}
\subsection{Large Separation Asymptotics: ``Intermediate Regime''}
Taking the limit $\eta \rightarrow \pm \infty$ in Eq.~(\ref{P-large-separation}) and  using Eqs.~(\ref{eta_R}, \ref{relation-eta-eta_R-1}), we obtain the large separation 
asymptotics for the interaction between two infinitely charged plates: 
\bea
P_{\rm net} = 
\left\{ 
 \begin{array}{ll} 
 \frac{8 T }{\pi \ell_{DB}^2 \lambda_{Bj}} e^{-L/\ell_{\rm DB}}, 
&
\pm \infty \,\, \mbox{plates in} \,\, q:-q. 
\vspace{3mm} \\
 \frac{18 T}{\pi \ell_{DB}^2 \lambda_{Bj}}  e^{-L/\ell_{DB}}, 
 & 
+ \infty  \, \, {\rm plates \, \, } 2q:-q;
\vspace{3mm} \\
 \frac{18 (2 - \sqrt{3})^2T}{\pi \ell_{DB}^2 \lambda_{Bj}}  e^{-L/\ell_{DB}},  
&
 - \infty  \, \, {\rm plates\, \, } 2q:-q. 
\end{array}
 \right.
\eea
Note the asymmetry between positively and negatively charged plates in asymmetric electrolytes.  
 
\subsection{Small Separation Asymptotics: ``Gouy-Chapman Regime''}

When $L \ll \ell_{DB}$,  the potential at the middle is high, so that the co-ion density is negligibly small between two plates.  The corresponding term is exponentially small and can be dropped from the PB equation.  The first integral then approximately reduces to 
\bea
\alpha = 
\left\{ \begin{array}{ll}
 \frac{1}{2} e^{\Psi} - \frac{1}{2 } (\partial_z \Psi)^2, 
&   \pm \infty \,\, \mbox{plates} \,\,q:-q 
\vspace{3mm}\\
 \frac{1}{3} e^{\Psi} - \frac{1}{2 } (\partial_z \Psi)^2, 
&  +\infty \,\, \mbox{plates}  \,\,2q:-q 
\vspace{3mm}\\
 \frac{1}{6} e^{-2\Psi} - \frac{1}{2 } (\partial_z \Psi)^2, 
&- \infty \,\, \mbox{plates}  \,\,2q:-q
\end{array}
\right.,
\eea
which can be integrated easily to yield:
\be
\Psi(z) =  \left\{ \begin{array}{ll}
2 \, \log \frac{2\,K}{\cos K z}, 
&   \infty \,\, \mbox{plates} \,\,q:-q 
\vspace{3mm}\\
2 \, \log \frac{2\,K}{\cos K z} + \log \frac{3}{2}, 
&  +\infty \,\, \mbox{plates} \,\,2q:-q 
\vspace{3mm}\\
- \log \frac{4\,K}{\cos 2K z} - \frac{1}{2}  \log \frac{3}{4},
& - \infty \,\, \mbox{plates} \,\,2q:-q
\end{array}
\right., 
\vspace{3mm}\\
\ee 
where $K = \sqrt{\alpha/2}$.  $\Psi(z)$ diverges at $z = \pm \pi /2 K$ for the first two cases, and at $ \pm \pi/ 4 K $ for the last case, therefore the distance between two infinitely charged plates is given by $L = {\pi}/{ K} = \pi \sqrt{{2}/{\alpha}}$ for the first two cases and $L = \pi/\sqrt{2 \alpha}$ for the last case.  $\alpha$ as a function of $L$ is then given by
\be
\alpha =  \left\{ \begin{array}{ll}
 2 \pi^2/L^2, 
&  \pm \infty \,\, \mbox{plates} \,\,q:-q 
\vspace{3mm}\\
2 \pi^2/L^2, 
&  +\infty \,\, \mbox{plates} \,\,2q:-q 
\vspace{3mm}\\
\pi^2/2\,L^2, 
& - \infty \,\, \mbox{plates} \,\,2q:-q
\end{array}
\right..
\ee 
Using Eq.~(\ref{P_net}) and Eq.~(\ref{P-alpha-asym}), the pressure, in physical unit, is then given by 
\be
P =  \left\{ \begin{array}{ll}
 \frac{2 \pi^2 \epsilon T}{q^2 L^2}, 
&  \pm \infty \,\, \mbox{plates} \,\,q:-q 
\vspace{3mm}\\
 \frac{2 \pi^2 \epsilon T}{q^2 L^2}, 
 &  +\infty \,\, \mbox{plates} \,\,2q:-q 
\vspace{3mm}\\
 \frac{\pi^2 \epsilon T}{2 q^2 L^2}, 
 & - \infty \,\, \mbox{plates} \,\,2q:-q
\end{array}
\right.,
\ee 
which is independent of the co-ions and the surface charge density. 

\subsection{Arbitrary Separation}
Inside a symmetric electrolyte,  Eq.~(\ref{Psi-p}) gives the potential between two infinitely charged plates located at $z = 0, 2 \omega_1$, with $\omega_1$ given by Eq.~(\ref{omega-1-alpha}).  Replacing $\omega_1$ with $L/2$ in  Eq.~(\ref{omega-1-alpha}), we find the relation between $L$ and $\alpha$: 
\be
L  = 2\,\Upsilon_{\rm WHP} \left( 
\frac{\alpha ^2}{3} - \frac{1}{4},
\frac{\alpha ^3}{27} - \frac{\alpha}{24}
\right), 
\quad \pm \infty \,\,\mbox{plates} \,\,
q:-q.   
\label{alpha-L-infinite}
\ee  

\begin{figure}
\begin{center}
\includegraphics[width=8.5cm]{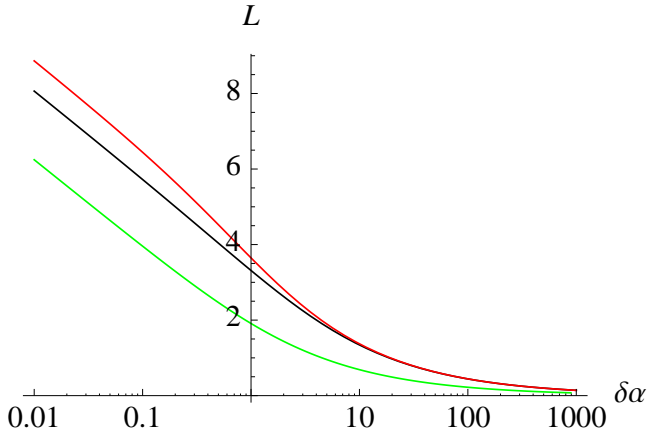}
\caption{ Log-linear plots of $L$ v.s. $\delta \alpha$ for two infinitely charged plates.  
Black: $q:-q$ symmetric electrolyte, as given by Eq.~(\ref{alpha-L-infinite}).  
Red: $\eta = +\infty$, $2q:-q$ electrolyte, see Eq.~(\ref{alpha-L-max-asym-1}).
Green: $\eta = - \infty$, $2q:-q$ electrolyte, see Eq.~(\ref{alpha-L-max-asym-2}).  
}
\label{period-A-plot}
\end{center}
\end{figure}

Inside a $2q:-q$ asymmetric electrolyte, the potential Eq.~(\ref{Psi-asym}) is always positive and has logarithmic singularities at $z = 0, 2 \omega_1$ as long as $\alpha >1/2$.  It therefore gives the potential between two $\eta = + \infty$ plates located at $0, 2 \omega_1(\alpha)$, with two invariants given by Eq.~(\ref{g-A-asym}).   The relation between $L$ and $\alpha$ is then given by
\be
L  = 2\,\Upsilon_{\rm WHP} \left( 
\frac{\alpha ^2}{3} ,
\frac{\alpha ^3}{27} - \frac{1}{108} 
\right), 
\quad  +\infty \,\, \mbox{plates} \,\,
 2q:-q . 
\label{alpha-L-max-asym-1}
\ee

\begin{figure}
\begin{center}
\includegraphics[width=7cm]{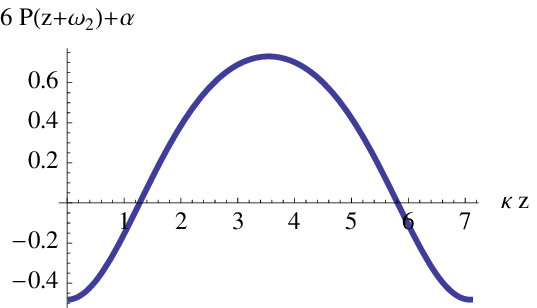} 
\\
\vspace{5mm}
\includegraphics[width=8cm]{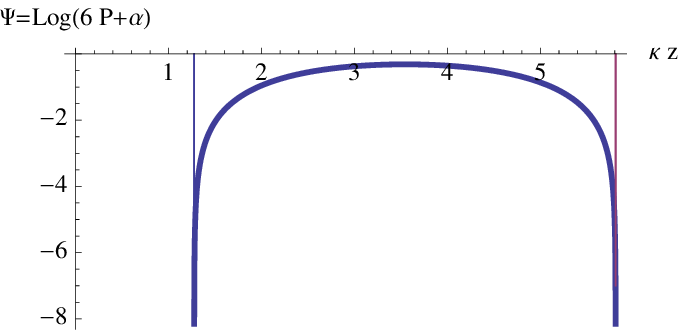}
\caption{ Up:   The function $6 \, \wp(z+ \omega_2; g_2,g_3) + \alpha$ has two zeros $(\zeta_1, \zeta_2) $ within one period. Between these two zeros, $0 < 6\,\wp +\alpha < 1$.   For the case shown in this plot, $\alpha = 5/9$, $(2\, \omega_1, 2 \omega_2) = (7.084, 6.225 i)$, $(\zeta_1, \zeta_2) = (1.275,5.810)$.  
\quad Down: For $\zeta_1 < z < \zeta_2$, $\Psi = \log (6 \, \wp + \alpha) <0 $ describes the potential between two negatively charged plates. The potential has two logarithmic singularities at $(\zeta_1,\zeta_2)$. }
\label{wp-asym-negative}
\end{center}
\end{figure}

The potential between two negatively charged plates inside a $2q:-q$ asymmetric electrolyte is also given by Eq.~(\ref{Psi-asym}), but with $z$ not lying on the real axis.  In fact, we find that for $\alpha >1/2$, $ 6\, \wp(z + \omega_2; g_2, g_3) + \alpha$ (with $z$ a real variable) has two zeros $\zeta_1, \zeta_2$ within one period. Between these two zeros, $0 < 6\,\wp +\alpha < 1$.   Note that $\omega_2$ is always purely imaginary for $\alpha > 1/2$.  In Fig.~\ref{wp-asym-negative}  we illustrate the function $6 \, \wp(z+ \omega_2; g_2,g_3) + \alpha$ with $\alpha = 5/9$, $(2\, \omega_1, 2 \omega_2) = (7.084, 6.225 i)$.The function has two zeros $(\zeta_1, \zeta_2)  = (1.275,5.810) $ within one period.   Since $\wp$ is an even function of $z$, we have $\zeta_2 = 2 \omega_1 - \zeta_1$.  As is also shown in Fig.~\ref{wp-asym-negative}, the potential $\Psi = \log (6 \, \wp + \alpha) $ has two logarithmic singularities at $(\zeta_1,\zeta_2)$; it is negative between these two singuarlties.  $\Psi $ therefore describes the potential between two $\eta = - \infty$ plates located at $\zeta_1,\zeta_2$ respectively.  For the case shown in Fig.~\ref{wp-asym-negative}, the distance between two plates is $\zeta_2 - \zeta_1 = 2 \omega_1 - 2 \zeta_1 = 4.535$.  

For arbitrary $\alpha$, there is no analytic expression for the two zeros $(\zeta_1,\zeta_2)$.   Nevertheless, we can alway find them numerically using Mathematica by solving the following equations:  
\bea
&& 6\, \wp\left( 
 \zeta_{i}(\alpha) + \omega_2(\alpha); 
\frac{\alpha ^2}{3} ,
\frac{\alpha ^3}{27} - \frac{1}{108} 
 \right) + \alpha 
= 0, \\ 
&& \quad 0< \zeta_1 (\alpha ) 
< \zeta_2 (\alpha ) < 2 \omega_1(\alpha). 
\nonumber
\eea
The relation between $L$ and $\alpha$ for a pair of negative infinitely charged plates is given by 
\bea
L =  2\,\Upsilon_{\rm WHP} \left( 
\frac{\alpha ^2}{3} ,
\frac{\alpha ^3}{27} - \frac{1}{108} 
\right)
- 2 \zeta_1(\alpha), 
\nonumber\\
\quad  - \infty \,\, \mbox{plates} \,\,
 2q:-q .  
\label{alpha-L-max-asym-2}
\eea

The relations Eqs.~(\ref{alpha-L-infinite}, \ref{alpha-L-max-asym-2}, \ref{alpha-L-max-asym-2}) are plotted in Fig.~\ref{period-A-plot}.

\section{Two Weakly Charged Plates}
\label{sec:two-plates-weak}
\subsection{Small Separation Asymptotics: ``Ideal Gas Regime'' and Beyond}
Here we focus on the case of symmetric electrolytes only.  If $L  \ll 1$, and $\eta \ll 1$, the potential Eq.~(\ref{Psi-p}) between plates can be expanded into Taylor series around $z = \omega_1$ up to $\delta z^2$: 
\be
\Psi(z) = \Psi(\omega_1) + \frac{1}{2} \Psi''(\omega_1) \delta z^2
+ O(\delta z^4). 
\ee
The condition under which the expansion is quantitatively accurate will be clear below.  $\Psi''(\omega_1)$ can be calculated using Eqs.~(\ref{Psi-p}), Eq.~(\ref{Elliptic-ODE}), and Eq.~(\ref{wp-omega1}):
\be
\Psi''(\omega_1) = \sqrt{\alpha^2 - 1}.
\ee
The Neumann boundary condition Eq.~(\ref{Neumann-BC}) (evaluated at $\delta z = L/2$) then gives:
\be
\alpha = \sqrt{1 + \frac{4 \eta^2}{L^2}} ,
\quad \quad q:-q .
\label{alpha-L-asym}
\ee
The net interaction between plates can be calculated using Eq.~(\ref{P_net}):
\be
P_{\rm net} = \frac{T}{4 \pi \ell_{DB}^2 \lambda_{Bj}} 
\left( \sqrt{1 + \left( \frac{2 \ell_{DB}^2 } {\ell_{GC} L } \right) ^2} -1 \right),
\quad \quad q:-q . 
\ee 
The Taylor expansion works if $\Psi''(\omega_1) L/2 \ll 1$.  Using the solution Eq.~(\ref{alpha-L-asym}) we find that this condition translates into 
$\eta \, L  \ll 1$, or in physical units, 
\be
L \ll \ell_{GC}, \ell_{DB},
\ee
which is the top right quarter in Fig.~\ref{four-regimes}.  This result has not been derived previously.  

This quarter can be divided into two regimes: if $\ell_{DB}^2/\ell_{GC} L  \gg 1$, we have $\alpha \approx 2 \eta /L$, which is the {\em ideal gas regime} \cite{Andelman-PBequation}, shown in Fig.~\ref{four-regimes}.  The net interaction in physical units is then approximately given by
\bea
P \approx \frac{T}{2 \pi \lambda \ell_{GC}L}
= \frac{2 T \sigma}{qL}.
\eea
In this regime, all co-ions are excluded from the region between plates by the strong Coulomb repulsion.  The counter-ions, on the other hand, behaves as an ideal gas with number density $2 \sigma /(q L)$.  If $\ell_{DB}^2/\ell_{GC} L  \ll 1$, which is part of the {\em Debye-H\"{u}ckel} regime where the separation is much shorter than the Debye length $L \ll \ell_{DB}$, $\delta \alpha \approx 2 \eta^2 /L^2$.   The net interaction in physical unit is then give by 
\bea
P_{\rm net} =  \frac{T \ell_{DB}^2}{2 \pi \lambda_{Bj} \ell_{GC}^2 L^2}.  
\eea
Both sub-regimes are discussed in the review by Andelman \cite{Andelman-PBequation}.  

Similar analyses can also be carried out for the cases of asymmetric electrolytes.  The results however are rather complicated, and not particularly illuminating.  We therefore only present asymptotic behaviors in the ideal gas regime, where $L \ll \ell_{GC}$ and $\ell_{DB}^2/\ell_{GC} L \gg 1$:
\bea
P_{\rm net} = \left\{
\begin{array}{ll}
\frac{2T \sigma}{qL} 
& + \mbox{plates }\,\, 2q:-q,
\vspace{3mm}\\
\frac{T \sigma}{qL} 
&  - \mbox{plates }\,\, 2q:-q.  
\end{array}
\right.
\eea 
The asymptotics of potential in the Debye-H\"{u}ckel regime will  be discussed below.  


\subsection{Large Separation Asymptotics: ``Debye-H\"{u}ckel'' Regime}
Now consider two weakly charged plates separated by a distance that is not too small, so that the potential is small everywhere $\Psi \ll 1$.  The Poisson-Boltzmann equation can then be linearized.  This is the Debye-H\"{u}ckel regime, where nonlinearities become irrelevant and {\em different electrolytes yield the same interaction and potential profile,} as long as they have the same Debye length and temperature.   
The linearized PB can be easily solved to yield the electrostatic potential:
\bea
\Psi(z) = \eta \frac{\cosh z}{\sinh L/2},
\eea
where two plates are located at $\pm L/2$ respectively.  $\alpha$ is given by
\bea
\delta \alpha = \frac{\eta^2}{2 \sinh^2(L/2) }.  
\eea
Using Eq.~(\ref{P_net}) and Eq.~(\ref{P-alpha-asym}), the net interaction between two plates is
\be
P_{\rm net} 
= \frac{T \eta^2 }{8 \pi \ell_{DB}^2\lambda_{Bj}} 
\sinh^{-2}\left(\frac{L}{2 \ell_{DB}}\right).   
\label{P-DH}
\ee 
The far field asymptotics is 
\be
P_{\rm net} = \frac{T}{2 \pi \lambda_{Bj} \ell_{GC}^2} 
e^{-L/\ell_{DB}},
\quad L \gg \ell_{DB}.  \label{DH-far-field}
\ee
Note that Eq.~(\ref{P-large-separation}) can be obtained from Eq.~(\ref{DH-far-field}) by replacing $\eta$ with $\eta_R$. 

\section{Two Strongly Charged Plates, and the General Case}
\label{sec:general-case}

In the general case of finite surface charge density and arbitrary separation between the plates, we need to determine two parameters $\alpha, z_0$, for a given $\eta$ and $L$.  For this purpose we need two equations.  Note that $z_0$ by definition is the distance from the plate to the nearest singularity of $\wp(z)$ $z = 0$, see the plot of $\Psi$ in Fig.~\ref{potential-plot}.  Since two plates are located at $z_0, 2 \omega_1 - z_0$ and are separated by $L$, we obtain the first equation:
\be
L = 2 \omega_1(\alpha) - 2 z_0,
\label{z0A-1}
\ee
where $\omega_1(\alpha)$ is given by Eq.~(\ref{omega-1-alpha}) for the case of symmetric electrolyte and by Eq.~(\ref{omega-1-alpha-asym}) for the case of $2q:-q$ asymmetric electrolyte.  The other equation is obtained using the Neumann boundary condition Eq.~(\ref{Neumann-BC}), which can be transformed into 
\begin{subequations}
\label{z0A-2-0} 
\bea
&& - \wp'(z_0) 
=  \eta \left[  \wp(z_0) + \frac{\alpha}{6}  \right],
\quad
q:-q, \,\, \mbox{or}  \,\,  \eta>0 \,\, 2q:-q  , 
\nonumber\\
\label{z0A-2-0-1} \\
&& - \wp'(z_0+ \omega_2) 
=  \eta \left[  \wp(z_0+ \omega_2) + \frac{\alpha}{6}  \right], 
\quad
 \eta<0 \,\, 2q:-q ,
\nonumber\\
\label{z0A-2-0-2}
\eea
\end{subequations}
after using of Eq.~(\ref{Psi-p}) or Eq.~(\ref{Psi-asym}).  Note that both $\wp$ and $\wp'$ depend on $\alpha$ through two invariants $g_2(\alpha),g_3(\alpha)$, see Eq.~(\ref{g-A}) or Eq.~(\ref{g-A-asym}), while the prime in $\wp'$ is with respect to $z$.  We are unable to obtain analytic solution to Eqs.~(\ref{z0A-1}, \ref{z0A-2-0}) in the general case.  Nevertheless, simple result can be obtained if $\eta \gg 1$.  

Let us first consider the first two cases given by Eq.~(\ref{z0A-2-0-1}).  Since $z_0 = 0$ for $\eta = \infty$ (see Fig.~\ref{potential-plot}), we expect $z_0 \ll 1$ for $\eta \gg 1$. Near $z_0=0$, the elliptic function in Eq.~(\ref{z0A-2-0}) can be expanded in terms of $z_0$ as
\be
\wp(z_0) = \frac{1}{z_0^2} + O(z_0^2).  
\ee
Note that the leading order term is independent of $\alpha$.  Hence Eq.~(\ref{z0A-2-0-1}) reduces to 
\bea
&& - \frac{2}{z_0^3} + \eta \frac{1}{z_0^2} + \frac{1}{6} \eta \alpha = 0, 
\nonumber\\
&& \eta = \frac{2}{z_0} (1+ z_0^2 \alpha/6)^{-1} 
\approx \frac{2}{z_0}.  \nonumber
\eea
Plugging it back into Eq.~(\ref{z0A-1}) we find that $\eta \gg 1$,  
\be
 2 \omega_1(\alpha) = L + 2 z_0 
 \approx L + \frac{4}{\eta}, 
\quad
q:-q, \,\, \mbox{or}  \,\, \eta>0 \,\, 2q:-q. 
\ee
Now consider Eq.~(\ref{z0A-2-0-2}).  Since the right hand side vanishes linearly at $z_0= \zeta_1$, see the discussion after Eq.~(\ref{alpha-L-max-asym-1}), we can expand it around $\zeta_1$:
\be
\wp (z_0+\omega_1) + \frac{1}{6} \alpha
= c_1 \, (z_0 - \zeta_1) + \cdots. 
\ee
Substituting this back into Eq.~(\ref{z0A-2-0-2}) we find 
\be
z_0 \approx \zeta_1 - \frac{1}{\eta}. 
\ee
Combining this with Eq.~(\ref{z0A-1})  we obtain 
\be
 2 \omega_1(\alpha) =
 L + 2 \zeta_1 - \frac{2}{\eta}, 
\quad \eta<0 \,\, 2q:-q. 
\ee

Combining the above results with Eq.~(\ref{omega-1-alpha}) or  Eq.~(\ref{omega-1-alpha-asym}) we find the following relations for two strongly charged plates with $\eta \gg 1$: 
\be
\begin{array}{ll}
 2 \,\Upsilon_{\rm WHP} \left( 
\frac{\alpha ^2}{3} - \frac{1}{4},
\frac{\alpha ^3}{27} - \frac{\alpha}{24}
\right)
= L + \frac{4}{\eta},  
& \pm \mbox {plates } q:-q 
\vspace{3mm} \\
 2 \,\Upsilon_{\rm WHP} \left( 
\frac{\alpha ^2}{3},
\frac{\alpha ^3}{27} - \frac{1}{108}
\right)
= L + \frac{4}{\eta}, 
&
+ \mbox{plates } 2q:-q 
\vspace{3mm} \\
2 \,\Upsilon_{\rm WHP} \left( 
\frac{\alpha ^2}{3},
\frac{\alpha ^3}{27} - \frac{1}{108}
\right)
= L + 2 \zeta_1(\alpha) - \frac{2}{\eta},  
&
- \mbox{plates } 2q:-q 
\end{array}. 
\label{alpha-strong-charge}
\ee
These result hold in the whole strongly charged regime where the Gouy-Chapman length $\ell_{GC}$ is much shorter than the Debye length $\ell_{DB}$.  This includes many examples in biological physics.  For a given surface charge density $\eta$, the $\alpha-L$ curve corresponding to Eqs.~(\ref{alpha-strong-charge}) can be obtained by a rigid shift of the curves for Eqs.~(\ref{alpha-L-infinite}, \ref{alpha-L-max-asym-2}, \ref{alpha-L-max-asym-2}) corresponding to infinitely charged plates along the $L$ axis.  

\begin{figure}
\begin{center}
\includegraphics[width=9cm]{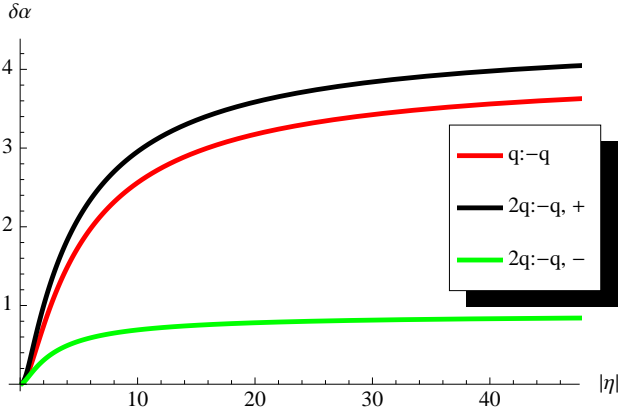}
\caption{The net  pressure $(\alpha - \alpha_0)$ as a function of the surface charge density $|\eta|$ for given plate separation $L = 2$, obtained by solving Eqs.~(\ref{z0A-1}, \ref{z0A-2-0})  All quantities are dimensionless.  Red: charged plates in a symmetric electrolyte.  Black: positively charged plates in a $2q:-q$ asymmetric electrolyte.  Green: negatively charged plates in a $2q:-q$ electrolyte.  Note how multi-valence counter-ions can dramatically reduce the pressure between two plates.  }
\label{alpha-eta}
\end{center}
\vspace{-3mm}
\end{figure}

Finally for the general case of arbitrary surface charge density, we can use Eq.~(\ref{z0A-1}) and Eqs.~(\ref{z0A-2-0-1}, \ref{z0A-2-0-2}) to numerically solve for $\alpha$ with given 
$L, \eta$.  For example, for a given distance between plates $L = 2 \ell_{DB}$, we plot the reduced net interaction $\delta \alpha$ as a function of the absolute value of the surface charge density $|\eta|$ in Fig.~\ref{alpha-eta} for all three cases.  One can explicitly see how the  interactions saturate as the surface charge density $|\eta|$ approaches infinity.  One can also see how divalent counter-ions dramatically reduces the repulsion between two charged plates, comparing with the case of monovalent counter-ions.  With slightly more efforts, we can also plot $\alpha $ as a function of separation $L$ for given surface charge density $\eta$.  We shall however not discuss this in detail here.  

\section{Conclusion and Acknowledgement}
\label{sec:conclusion} 
We have discussed in great detail the exact solution of the Poisson-Boltzmann equation in the two-plate geometry, both in $q:-q$ symmetric and in $2q:-q$ asymmetric electrolytes.  The Weierstrass elliptic representation of the potential has numerous advantages: 1) it yields simple analytic results for the one-plate problem in $2q:-q$ asymmetric electrolytes; 2) it yields novel and exact asymptotic results in various regimes, both in symmetric electrolytes and in $2q:-q$ asymmetric electrolytes; 3) it allows us to see explicitly how multi-valence counter-ions significantly reduce the repulsion between two likely charged plates, as well as how the renormalized surface charge density depends on the bare surface charge density for the one plate problem. 

Our exact solutions of PB may also tell us what to be expected beyond the Poisson-Boltzmann theory.   There are at least three important issues that are ignored by the PB.  Firstly, when the surface charge density is sufficiently high, the counter-ion density near the plates may become so high that they may crystallize on the plate surface.  This possibility has been extensively explored by many authors, but has not been completely clarified.  Secondly, if the ions in the electrolyte are of multiple valence, the Bjerrum length can be much longer than the ion diameter, and most ions form neutral bound pairs.   The response of these pairs to the external charged objects is a serious problem, but has not been explored so far.  Finally, there may be important short range chemical interactions between particular types of ions and the charged surfaces.  These interactions in principle must exist in reality, and can be easily taken into account in a field theoretic formalism of the problem.  Study of the latter two issues will be the major mission of our future works.  

The author thanks Leo Radzihovsky,  Andy Lau, Hongru Ma, Erik Luijten, Michael Brenner, and Anatoly Kolomeisky for helpful discussions on the general field of electrolytes. 

\appendix

\section{Derivation of Poisson-Boltzmann Equation}
\label{app:derivation-PB}
There are many different derivations of the Poisson-Boltzmann equation in the literature.  To make this work self-contained, we present two different derivations  based on the variational principle.  The equations derived from these two approaches are however slightly different.  Our derivations have the merit of  explicitly treating the spatial variation of dielectric constant and the electrostatic boundary conditions. 

\subsection{Derivation From the Grand Canonical Ensemble}

Let us start with a symmetric electrolyte with some charged dielectric solids fixed in the space.   The ions move inside the solvent but are not allowed to penetrate into the solids.  The grand canonical partition function for the system of ions can be mapped into the generating functional of the sine-Gordon field theory:  
\bea
Z &=& C^{-1} \, \int D \varphi \, e^{-\beta H[\varphi]},
\label{model-0}\\
H[\varphi] &=& \int dx \left[ 
\frac{1}{2} \epsilon(x) (\nabla \varphi)^2
- 2 T\, n \,(x) \, \cos \beta q \varphi 
\right]
\nonumber\\
&+& i \oint_{\partial \Omega} \varphi \sigma dA, 
\nonumber
\eea
where $\beta = 1/T$, $q$ is the charge of all ions, while  $C$ is a normalization constant that is irrelevant to our purpose.   $\Omega$ is the spatial region occupied by the dielectric solids, while $\partial \Omega$ is the surface of dielectrics.   The functions $\epsilon(x), n \,(x)$ are given by  
\be
\epsilon(x) = \left\{ 
\begin{array}{ll}
\epsilon_s, & x \in \Omega \\
\epsilon_l, & x \notin \Omega
\end{array}
\right. ,
\quad
\quad
n \,(x)= 
 \left\{ 
\begin{array}{ll}
 0, 
& x \in \Omega \\
n \, , 
& x \notin \Omega
\end{array}
\right. .
\ee
At the saddle point level, $n \,$ can be understood as the density of each specie of ions, so that the total ion density is $2 n \,$.  $\sigma$ is the surface charge density prescribed externally on the surface of dielectrics, assumed to be constant through out this work.  Because of the imaginary boundary term, the saddle point value of the order parameter $\varphi$ (and its average as well) is purely imaginary.  The average of $i \varphi$ has the physical significance of  the average electrostatic potential.   A detailed derivation will be published elsewhere.

Minimization of the action over $\Phi = i \varphi$, with the boundary terms properly taken into account, leads to the Sine-Gordon equation 
\bea
&& - \epsilon_l \Delta \Phi + 2 n \, q \sinh \beta q \Phi = 0, 
\quad x \in \Omega
\label{Saddle-solvent}
\\
&& - \epsilon_s \Delta \Phi = 0, 
\quad x \notin \Omega 
\label{Saddle--solids}
\eea 
together with the standard electrostatic  boundary condition on the dielectric interfaces:
\be
\left.\epsilon_l \frac{\partial \Phi}{\partial n} \right|_{\partial \Omega, l}
- \left. \epsilon_s \frac{\partial \Phi}{\partial n} \right|_{\partial \Omega, s}
- \sigma = 0,
\label{Boundary-condition}
\ee
where the unit normal vector $\hat{n}$ goes from the solvent into the solid dielectrics.  Note that the boundary condition Eq.~(\ref{Boundary-condition}) is not satisfied by  a generic configuration of the field $\varphi$, but only by the saddle point configuration.   Now for aquarius electrolytes  $\epsilon_l \approx 80 \gg \epsilon_s \sim 1$, we can ignore the term proportional to $\epsilon_s$ in Eq.~(\ref{Boundary-condition}).   This leads to the Neumann boundary condition:
\be
\left. \epsilon_l \frac{\partial \Phi}
{\partial n} \right|_{\partial \Omega, l}
 = \sigma, 
 \label{Neumann-BC-1}
\ee
hence the region inside the dielectric solids is decoupled from the region occupied by the electrolyte.  We shall take this approximation in this work.

\subsection{Derivation from Canonical Ensemble}
We shall present a slightly different derivation of PB from the canonical ensemble.  First consider the simple case of uncharged solid dielectrics inside a solvent with no ion.  A unit point charge\footnote{In reality ions have finite radius and there is dispersion force between them and the macroscopic charged objects.  We shall ignore this force in this work.  } at $y$ inside the solvent induces surface charges on the dielectric boundaries.  Inside the solvent, the electrostatic potential, i.e. the Green's function of the corresponding electrostatic problem, can be written as 
\be
G(x,y) = \frac{1}{4\pi\epsilon_l |x-y|} + \chi(x,y) ,
\label{G-chi}
\ee
where the first term is understood as the potential generated by the source charge, while the second term $\chi(x,y)$ is the potential by all the induced surface charges, given that there is a unit source charge at $y$.   One can easily show that the interaction energy between the point charge $q$ and image charges it induces is given by
\be
W(y) = \frac{1}{2}q^2 \chi(y,y) \label{W-chi}. 
\ee
It can be proven that the Green's function satisfies the following PDE:
\be
- \nabla \cdot \epsilon (x) \nabla G(x,y) = \delta(x-y).  
\label{Green-def}
 \ee

Now consider the problem of charged dielectric solids inside electrolyte. The total energy of all mobile ions (with charges $q_i$ and location $r_i$) is given by 
\be
H = \frac{1}{2} \sum_{i \neq j} q_i q_j G(r_i,r_j) + \frac{1}{2} \sum_{i} q_i^2 \phi_1(r_i)
+ \sum_{i} q_i \phi_{\rm ext}(r_i),
\ee
where the sums are over all the mobile ions, and 
\be
\phi_1(y) = \chi(y,y)
\ee
is the image charge potential acting on the ion at $y$, while $\phi_{\rm ext}$ is the potential produced by all fixed charges on the surface of dielectrics.  We want to calculate the canonical partition function \footnote{Strictly speaking, the integrals  in the partition function are all divergent, because the Boltzmann factor increases without bound when two opposite point charges approach each other.  These divergences have to be regularized by introducing some hard core radius for the ions, within which other ions can not penetrate.  At the level of mean field theory however this issue is irrelevant.  }
\be
Z = \int \prod_i dr_i \, e^{- \beta H} ,
\ee
where the integrals are over the positions of all mobile ions, restricted inside the solvent, i.e. not allowed to penetrate into the solids.   We consider the following variational probability distribution function (pdf): 
\be
P[ \{ r\} ] = \prod_{\alpha = 1}^{N_+} p_+(x_{\alpha}) 
\prod_{\beta = 1}^{N_-} p_-(y_{\beta}),
\ee
where we use coordinates $x_{\alpha}$ and $y_{\beta}$ for positive and negative ions.  Since all spatial variables are independent to each other in this pdf, we  are ignoring all the spatial correlations between different ions, a hallmark of all mean field theories.   

We calculate the variational free energy
\bea
F_{\rm var} &=& \langle H \rangle_{\rm var} - T \, S_{\rm var}
\\
&=& \int \prod_i dr_i  \,  \left( P[ \{ r\} ]  \,   H
+ T  P[ \{ r \} ] \log P[ \{ r\} ]  \right).  
\nonumber
\eea
This variational free energy then provides upper bound to the exact free energy, according to Feynman's variational principle.   

The variational free energy $F_{\rm var}$ is more conveniently expressed in terms of  ion number densities 
\be
n_{\pm} (r) = N_{\pm} p_{\pm}(r).  
\ee
After some calculation and ignoring terms smaller by orders of $N^{-1}$, we find 
\bea
F_{\rm var}  &=& 
\frac{q^2}{2} \int_1 \int_2 [ n_+(1) - n_-(1) ] G(1,2) [ n_+ (2) - n_-(2) ] 
\nonumber\\
&+&  q \int_ r \phi_{\rm ext}(r) [ n_+ (r) - n_-(r) ]
\nonumber\\
&+&  \frac{q^2}{2} \int_r \phi_1(r)  [ n_+ (r) + n_-(r) ], 
\nonumber\\
&+& T \int_r \left[ 
n_+(r) \log n_+(r) + n_-(r) \log n_-(r)
\right], \label{free-energy-var}
\eea
where symbols $1,2$ stand for $r_1, r_2$.  All the integrals are restricted outside the dielectrics.  Note that the image charge potential $\phi_1$ explicitly contributes to the free energy.  

Now define the averaged total potential $\Phi(r)$ by all the charges, including the fixed surface charges, the induced surface charges, as well as the mobile ions.  It is given by
\bea
\Phi (1) &=& q \int_{2} G(1,2) 
\left[ n_+(2) - n_-(2)\right] + \phi_{\rm ext} (1),
\nonumber\\
&=&  \int_{2} G(1,2) 
\rho(2) + \phi_{\rm ext} (1),
\label{phi-tot-def}
\eea
where
\be
\rho(r) = q [ n_+ (r) - n_-(r) ] 
\label{rho-n}
\ee
is the average mobile charge density.  Note that in Eq.~(\ref{phi-tot-def}) the total potential $\Phi(x)$ is defined in the whole space, including the region inside the dielectrics, even though the charge density $\rho(x)$ is nonvanishing only in the electrolyte.  Varying the free energy  Eq.~(\ref{free-energy-var}) over the number densities $n_{\pm}$ subject to the constraints of fixed total numbers of $\pm q$ ions, we find the saddle point equation as
\bea
n_{\pm}(r) = n \, \, \exp \left[  \mp \beta q \Phi(r) - \frac{1}{2} \beta q^2 \phi_1 (r) \right],
\label{n-varphi}
\eea
where $n \,$ serves as the Lagrange multiplier fixing the total number of ions.  As $r \rightarrow \infty$, both $\Phi$ and $\phi_1$ approach zero, hence $n_{\pm} \rightarrow n$.  Hence $n$ is bulk ion density for each specie.   Note how this saddle point equation explicitly depends on the image potential $\phi_1(r)$!  Using  Eq.~(\ref{n-varphi}), the average charge density $\rho(x)$ due to all mobile ions can be expressed as 
\bea
\rho(r) &=& q [ n_+ (r) - n_-(r) ] 
\label{rho-phi}\\
&=& - 2 n \, q
\exp \left[ -\frac{1}{2}  \beta q^2 \phi_1(r) \right]
\sinh [ \beta q \Phi(r)] 
\nonumber
\eea

Now let the operator $- \nabla \cdot \epsilon(x) \nabla$ acting on Eq.~(\ref{phi-tot-def}), and use Eq.~(\ref{Green-def}), as well as the fact that 
\be 
- \nabla \cdot \epsilon(x) \nabla   \phi_{\rm ext} 
=  \rho_{\rm ext} (x),
\ee
we find 
\bea
- \nabla \cdot \epsilon(x) \nabla \Phi (x) = \rho(x) +  \rho_{\rm ext} (x), 
\label{Phi-eqn}
\eea
where $\rho_{\rm ext}(x)$ is the externally prescribed surface charge density, which is nonzero only on the dielectric interface.  Let us now consider the regions inside the electrolyte and that inside the dielectrics separately.  For the region inside the electrolyte, $\epsilon(x) = \epsilon_l$, while $\rho(x)$ is given by Eq.~(\ref{rho-phi}) and $ \rho_{\rm ext} (x) = 0$, therefore the equation Eq.~(\ref{Phi-eqn}) reduces to 
\be
-  \epsilon_l \Delta  \Phi =- 2 n \, q
\exp \left[ -\frac{1}{2}  \beta q^2 \phi_1(r) \right]
\sinh [ \beta q \Phi(r)] , 
\quad x \notin \Omega,   
\label{Saddle-solvent-1}
\ee
which is slightly different from Eq.~(\ref{Saddle-solvent}).   For the region $\Omega$ inside the solids, $\epsilon(x) = \epsilon_s$, $\rho(x) = \rho_{\rm ext}(x) = 0$, and therefore the equation Eq.~(\ref{Phi-eqn}) reduces to the Poisson equation  Eq.~(\ref{Saddle--solids}). 
Finally, the singular surface charge density term in  Eq.~(\ref{Phi-eqn}) results in the standard dielectric boundary condition Eq.~(\ref{Boundary-condition}).  

Substituting Eq.~(\ref{n-varphi}) back into Eq.~(\ref{free-energy-var}), we find that, up to a trivial additive constant, the variational free energy reduces to 
\bea
F_{\rm var} &=& - \frac{q^2}{2} \int_1 \int_2 [ n_+(1) - n_-(1) ] G(1,2) [ n_+ (2) - n_-(2) ] 
\nonumber\\
&=& - \frac{1}{2} \int_r \rho(r) \left[ \Phi (r) - \phi_{\rm ext} (r) \right]
\eea

That there exists two different versions of the saddle point equation, Eq.~(\ref{Saddle-solvent}) and Eq.~(\ref{Saddle-solvent-1}) should not bother us.  
Even though both equations are derived from variational principles, the underlying physics is rather different.  In the canonical ensemble approach, we directly vary the spatial pdf of every ion, while in the grand canonical ensemble approach, we vary $\varphi$, which is an auxiliary field introduced in a lattice model.  A more important difference between two approaches is that Eq.~(\ref{free-energy-var})
is real, while the Sine-Gordon action Eq.~(\ref{Saddle-solvent}) is complex.  As a consequence, the canonical ensemble approach also provides an upper bound for the free energy, while the grand canonical ensemble approaches does not.   
Other than that, however, it is {\it a priori} not so clear which of these two equations is a better approximation.  Eq.~(\ref{Saddle-solvent}) nevertheless has the advantage of being simpler.  

Quantitatively the difference between these two MFTs becomes relevant only when the exponent $- \beta q^2 \phi_1(r)/2 $ in Eq.~(\ref{Saddle-solvent-1}) is large.   For the simple case of an ion near a flat dielectric boundary with $\epsilon_ s \ll \epsilon_l \approx 80$, we can estimate the exponent as $
-{q^2}/{8 \pi \epsilon d \, T},$ where $d$ is the distance from the ion to the plate.  For monovalent ion, this factor is becomes larger than unity when $d$ is smaller than half of the Bjerrum lengh, $0.35 nm$.  A few times of this distance away from the dielectric boundary, we can safely ignore the issue of image charge potential altogether.   Eq.~(\ref{Saddle-solvent}) and Eq.~(\ref{Saddle-solvent-1}) then becomes identical.

\section{Some General Results at Saddle Point Level}
\label{app:general_results}
In this section, we derive some general results about the PB theory in a symmetric electrolyte.  We first define three useful length scales \footnote{Our definition of Gouy-Chapman length is different from the usual by a multiplicative factor. }: 
\bea
\ell_{DB} &=& \kappa^{-1} 
= \sqrt{\frac{\epsilon_l}{2 \beta n \,q^2}} 
\quad \mbox{Debye screening length},
\\
\ell_{GC} &=& \epsilon/ q \beta \sigma
\quad \quad \quad \quad \quad \mbox{Gouy-Chapman length},
\\
\lambda_{Bj} &=& \frac{q^2}{4 \pi \epsilon_l\, T}
\quad \quad \quad\quad\quad \mbox{Bjerrum length}.
\eea
The following relation between $\ell_{DB}$ and $\lambda_{Bj}$ shall be useful below:
\be
8 \pi \ell_{DB}^2 \lambda_{Bj} n = 1. 
\label{l-lambda-relation}
\ee
We further define the dimensionless versions of the potential, spatial coordinates as well as the surface charge density:   
\bea
\psi &=& \beta q \varphi, \\
\rv &=& \sqrt{\frac{2 \beta n \,q^2}{\epsilon_l}} \xv = \kappa \, \xv, \\
\eta &=& \frac{ \sigma}{\sqrt{2 \epsilon T n \,}} 
=  \frac{\ell_{DB}}{\ell_{GC}}.  
\eea
We shall also define a dimensionless temperature 
\be
\hat{T}  = \frac{1}{2 \, n \, \, \ell_{DB}^3}
= {4 \pi} \frac{\lambda_{Bj}}{\ell_{DB}}, 
\label{That-def}
\ee
as well as a dimensionless Hamiltonian 
\be
\hat{H} = H \hat{T}/T. 
\ee 
  The dimensionless temperature $\hat{T}$ is small for diluted electrolytes.  For example, in the recent experiment on highly deionized water by Tata {\it et al} \cite{tata-boundpairs-2008}, $\ell_{DB} \approx 500 nm, \lambda_{Bj} \approx  0.7 nm$, and $\hat{T} \approx 1/57$; the system is therefore well in the low temperature limit.  The problem of dilute electrolytes is therefore simple from the field theory point of view.

In terms of all these dimensionless variables, the grand canonical partition function Eq.~(\ref{model-0}) can be written as 
\bea 
Z &=& C^{-1} \int D \psi \, e^{- \hat{H}[\psi] /\hat{T}}, 
\label{H-hat-psi} \\
\hat{H}[\psi] &=& \int dx \left[  \frac{1}{2}
(\nabla \psi)^2 - \cos \psi 
\right]
+ i \oint_{\partial \Omega}  \eta \,\psi(\rv) \, dA.  
\nonumber
\eea
 Varying $\hat{H}$ over $\Psi = i \psi$, and taking into account the translational symmetry in the plane of plates, we find the saddle point equation
 \be
- \partial_z^2 \Psi + \sinh \Psi = 0.  
\label{saddle-eqn}
\ee 
The boundary condition Eq.~(\ref{Neumann-BC-1}) reduces to (again in the limit $\epsilon_s/\epsilon_l \rightarrow 0$)
\be
\left. \frac{\partial \Psi}{\partial n} \right|_{\rm boundary} = \eta.  
\ee

The saddle point approximation of the grand potential is given by:  
\bea
{\mathcal G} &=& T \log Z \approx
- H[- i \Psi] = - \hat{H}[-i \Psi] T/\hat{T},
\nonumber\\
\\
\hat{H}[- i \Psi] 
&=& A \int dz \left[ - \frac{1}{2} (\partial_z \Psi)^2
- \cosh \Psi \right]
+ A \sum_{i = 1,2} \eta \, \Psi_0,
\nonumber\\
\label{MFT-free-energy}
\eea
where $A$ is the area of the plates, while $\Psi_0 $ is the electrostatic potential on each plate, which should be determined using the boundary condition Eq.~(\ref{saddle-eqn}).   The potential is constant on the plate because of the translational symmetry.  It is no longer constant for the problem of non-planar objects.  Rather curiously, the action Eq.~(\ref{MFT-free-energy}) at the saddle point has a {\em negative} gradient term.

Even though we will be primarily concerned with the mean field theory, for completeness, we show the quadratic part of fluctuation Hamiltonian around the saddle point.  Let $\varphi = - i \Psi + \phi$.   After some simple calculation, we find
\be
\delta_2 \hat{H}[\phi,\Psi]
 = \int dx \left[  
 \frac{1}{2} (\nabla \phi)^2
+ \frac{1}{2} \cosh \Psi(r) \, \phi^2
 \right].  
\ee
It is a quadratic theory with a spatially variable mass $\cosh \Psi(\rv)$.  Interestingly, the mass becomes exponentially large where the saddle point potential $\Psi(\rv)$ is large, which makes fluctuations quantitatively unimportant.  

We readily obtain the following first integral for PB equation Eq.~(\ref{saddle-eqn}):
\be
{\alpha} = \cosh \Psi (z)- \frac{1}{2} (\partial_z \Psi(z))^2.   
\label{first-integral-1}
\ee
Let us show that  $\alpha \geq 1$ corresponds to the case of two likely charged plates, while $\alpha \leq 1$ corresponds to  the case of two oppositely charged plates.  If two plates carry the same charge density, in the middle of two plates, we have $\partial_z \Psi (MP)= 0$ due to symmetry, where $MP$ stands for ``mid point''.  Hence 
\be
\alpha = \cosh \Psi(MP) \geq 1. \label{A-L}
\ee
On the other hand, if two plates carry equal but opposite charges, then $\Psi(MP) = 0$ due to symmetry while $\partial \Psi (MP) \neq 0$, hence 
\be
\alpha = \cosh 0 - \frac{1}{2} (\partial_z \Psi(MP))^2 \leq 1. 
\ee 
When two plates are infinitely separated, both $\Psi$ and $\partial_z \Psi$ vanish in the middle, therefore $\alpha = 1$, regardless of the signs of charge on the plates. 

For the case of two likely charged plates, in no where between two plates can the potential $\Psi$ vanish, for this would lead to $\alpha   < 1$,  which contradicts our earlier observation Eq.~(\ref{A-L}).   Hence the mean field electrostatic potential does not change sign.  The so-called ``charge inversion'' does not happen at the saddle point level. 

The total ion density at $z$ is given by
\be
n_{\rm tot} (z)
= n \left(  e^{ \beta q \Phi} +  e^{-\beta q \Phi} \right) 
= 2 \,n \, \cosh \Psi (z). 
\ee
Infinite away from the plates, i.e., in the bulk, we have 
\be
\Psi(\infty) = 0, \quad 
n_{\rm tot}(\infty) = 2 \, n,
\label{A-n0}
\ee
while in the middle of two plates, we have
\be
n_{\rm tot}(MP) = 2 \, n \cosh \Psi(MP).  
\ee
Hence 
\be
\alpha = \cosh \Psi(MP) = \frac{n_{\rm tot} (MP)}{n_{\rm tot}(\infty) } 
\label{alpha-Psi}
\ee
is the ratio between the ion density in the middle of two plates to that of the bulk.   The fact that $\Psi(MP)>0$ means that there is a surplus of ion density in the middle of two plates. 


To find the interaction between two plates, we further rescale the coordiate:
\be
z \rightarrow Z = z/L, 
\ee
so that the grand potential Eq.~(\ref{MFT-free-energy}) becomes 
\be
\frac{ \hat{T}{\mathcal G}_{MF}}{A \,T} = 
 \int_0^1dZ \left[ - \frac{1}{2L} (\partial_Z \Psi)^2
- L \, \cosh \Psi \right]
+ 2 \, \eta \, \Psi_0,
\ee
where $A$ is the surface area of both plates.  The saddle point solution $\Psi$ generically depends on $L$. 
When we vary the above expression over $L$, however, we find that all terms involving $\delta \Psi$, including the surface term, cancel at the saddle point.  Therefore the variation of the free energy as we change the plate separation $L$ is given by
\bea
\frac{\hat{T}}{A\,T} \delta {\mathcal G}_{MF} &=& 
\frac{\delta L}{L} \int_0^1 dZ \left( 
 \frac{1}{2L} (\partial_Z \Psi)^2
- L \, \cos \Psi 
\right) 
\nonumber\\
&\rightarrow& 
\frac{\delta L}{L} \int_0^L dz \left( 
 \frac{1}{2} (\partial_z \Psi)^2
-  \cosh \Psi 
\right)
\nonumber\\
&=& - \alpha\, \delta L,
\label{alpha-dG}
\eea
where we have transformed back to the usual coordinate $z$ in the second equality, and have used Eq.~(\ref{first-integral-1}).  Restoring the physical units, then, we find that the force per unit area between two charged plates is given by \footnote{Note that $P$ here is not the pressure of the whole electrolyte as a fluid. It is the pressure between two plates due to the existence of charges. } 
\bea 
P  \equiv
 -  \ell_{DB}^{-3} \frac{1}{A} 
 \frac{\partial {\mathcal G}_{MF}}{\partial L}
= \frac{T}{\hat{T}} \ell_{DB}^{-3} \alpha 
= \frac{T\, \alpha}{ 4 \pi \ell_{DB}^2 \lambda_{Bj}}.  
\label{P-osmotic-sym}
\eea
In the first equality, the factor $\ell_{DB}^{-3}$ is needed to restore the physical unit for $P$; in the second equality, used was Eq.~(\ref{alpha-dG}); the third, Eq.~(\ref{That-def}) and Eq.~(\ref{l-lambda-relation}).  The last equality is useful, because it involves only temperature and two fundamental length scales for the electrolyte, as well as a dimensionless parameter $\alpha$.  
 Using Eq.~(\ref{A-n0}) and Eq.~(\ref{alpha-Psi}), we can also write the pressure as 
\be
P = 2 n T \alpha = n_{\rm tot} (MP)T
\ee
Therefore, $P$ is essentially the osmotic pressure of the ionic gas at the middle of plates.  

The ionic gas outside the plates also exert pressure on them, therefore $P$ calculated above is the not net force per unit area between two plates.   To obtain the net interaction, let us introduce another fictitious plate to the left of the left plate with a distance that is much greater than $L$.  The electrostatic potential is given by the same solution, but with the constant $\alpha  \approx 1$.  The resulting pressure acting on the plate (pointing to the right ) is then given by $2 n T$.  Hence the net pressure acting on the left plate is 
\be
P_{\rm net} =  2 n T ( \alpha -1) = 2 n T \delta \alpha,  
\ee
which vanishes as the separation $L$ between two plates goes to infinity, as it should be.  The fact that $\alpha >1$ means the interaction between two likely charged plates is always {\em repulsive}.  Like charge attraction does not happen for plates geometry at the level of mean field theory.   The result has been proven for more general geometry by Neu some time ago \cite{Neu-prl-1999}.  Finally using Eq.~(\ref{l-lambda-relation}), we can also express the net interaction as 
\be
P_{\rm net} = \frac{T\,\delta \alpha}{ 4 \pi \ell_{DB}^2 \lambda_{Bj}}. 
\ee

\section{$2q:-q$ Asymmetric Electrolyte}
\label{app:asymmetric}
An outstanding advantage of the Weierstrass function representation is that it allows an explicit calculation for the case of $2:1$ asymmetric electrolytes.  
Assume positive ions carry charge of $2 q$, while negative ions carry charge $-q$.    To keep overcall charge neutrality, the bulk ion densities are $n /2$ and $n $ for positive and negative ions respectively. The Sine-Gordon field theory for asymmetric the electrolyte is given by the following:
\bea
Z &=& C^{-1} \, \int D \varphi \, e^{-\beta H[\varphi]},
\label{model-asymmetric}\\
H[\varphi] &=& \int dx \left[ 
\frac{1}{2} \epsilon (\nabla \varphi)^2
-  T\, n \, \left(  e^{ i \beta q \varphi }
+  \frac{1}{2} e^{ - 2 i \beta q \varphi }
\right)
\right]
\nonumber\\
&+& i \oint_{\partial \Omega} \varphi \sigma dA, 
\nonumber
\eea
For the more general case of $k q$ positive ions, $e^{-2 i \beta q \varphi }/2$ should be replaced by $e^{-k i \beta q \varphi}/ k$.

The Debye length for a $2q:-q$ electrolyte is given by 
\be
\ell_{DB} = \sqrt{\frac{\epsilon}{3 n \, \beta q^2}}, 
\label{Debye-asym-app}
\ee
with $n \,$ the bulk density of the $-q$ ions.  We define the Bjerrum length $\lambda_{Bj}$ and the Couy-Chapman length the same way as in Sec. \ref{app:general_results}.   The analogue of the identity Eq.~(\ref{l-lambda-relation}) becomes
\be
12 \pi \ell_{DB}^2 \lambda_{Bj} n = 1.  
\label{l-lambda-relation-asym}
\ee
We rescale the field $\varphi$,  the real space coordinates as well as the surface charge density in the following way:
\bea
\psi &=& \beta q \varphi, \\
\rv &=& \sqrt{\frac{3 \beta n \,q^2}{\epsilon_l}} \xv =  \frac{\xv}{ \ell_{DB}}, \\
\eta &=& \frac{ \sigma}{\sqrt{3 \epsilon T n \,}} 
=  \frac{\ell_{DB}}{\ell_{GC}},\\
\hat{T} &= & 4 \pi \frac{\lambda_{Bj}}{\ell_{DB}}
 = \frac{1}{3 n \ell_{DB}^3}. 
\label{rescaling-asym}
\eea

After the rescaling the partition function can be written as
\bea
Z &=& \int D \psi \, e^{-\hat{H}/\hat{T}},
\\
\hat{H} &=& \int dx \left[ 
\frac{1}{2} (\nabla \psi)^2 
- \frac{1}{3} e^{i \psi}
- \frac{1}{6} e^{-2i \psi}
\right] 
+ i \oint_{\partial \Omega}  \eta \,\psi \, dA.  
\nonumber
\eea
The saddle point is again purely imaginary, $\psi = - i \Psi$, with $\Psi$ the electrostatic potential at the mean field level.  The saddle point equation satisfied by $\Psi$ is given by
\be
-  \partial_z^2 \Psi + \frac{1}{3} e^{\Psi} - \frac{1}{3} e^{-2 \Psi} = 0, 
\quad 
\left. \frac{\partial \Psi}{\partial {n}} \right|_{\rm plates}= \eta. 
\label{saddle-asym-app}
\ee
Linearization of this equation leads to Eq.~(\ref{PB-linear}), as it should be.  The first integral of this second order ODE can again be easily obtained:
\be
\alpha = - \frac{1}{2} (\partial_z \Psi)^2 + \frac{1}{3}e^{\Psi} + \frac{1}{6} e^{-2 \Psi}. 
\label{A-asymmetric-app}
\ee

Consider the case of two equally charged plates.  At the middle between two plates, $\partial_z \Psi (MP) = 0$, hence 
\be
\alpha =  \frac{1}{3}e^{\Psi(MP)} 
+ \frac{1}{6} e^{-2 \Psi(MP)}
= \frac{n_{\rm tot}(MP)}{3 n \,}
\geq 1/2, 
\label{alpha-Psi-asym}
\ee
where 
\be
n_{\rm tot}(MP) = n \, e^{\Psi(MP) } + \frac{1}{2} n \, e^{- 2 \Psi(MP) }
\ee
is the total ion density at the middle of plates.  The lower bound $\alpha = 1/2$ corresponds to the case $\Psi(MP) = 0$, where two plates are infinitely far away.  

Following a similar strategy as in Sec.~Sec. \ref{app:general_results}, we can prove the following result for the pressure between two plates: 
\be
P = 3 n T \alpha = \frac{T \alpha}{4 \pi \ell_{DB}^2 \lambda_{Bj}}.   
\ee
i.e. it is the osmotic pressure of the ion gas in the middle of plates.  The net interaction between two plates is then given by 
\be
P_{\rm net} = 3 n T \left( \alpha - {1}/{2} \right)
= \frac{T \delta \alpha}{4 \pi \ell_{DB}^2 \lambda_{Bj}}.  
\label{P-osmotic-asym}
\ee
Since $\alpha > 1/2$, two likely charged plates repel each other.  

In the bulk, $\Psi =  \partial_z \Psi = 0$, hence $\alpha = 1/2$.  Therefore the osmotic pressure of the ion gas in a $2q:-q$ electrolyte, according to Eq.~(\ref{P-osmotic-asym}) (with $\alpha$  set to $1/2$) is given by 
\be
P_{2q:-q}  = \frac{T}{8 \pi \ell_{DB}^2 \lambda_{Bj}}.  
\label{P-osmotic-asym-1}
\ee
By contrast, the osmotic pressure of the ion gas in a $q:-q$ symmetric electrolyte, according to Eq.~(\ref{P-osmotic-sym}) (with $\alpha$  set to unity), is given by
\be
P_{q:-q} = \frac{T}{4 \pi \ell_{DB}^2 \lambda_{Bj}},  
\label{P-osmotic-sym-1}
\ee
which is twice of $P_{2q:-q}$.  This result can be easily obtained by inspecting the expressions for Debye length in both cases, Eq.~(\ref{Debye-length-def}) and Eq.~(\ref{Debye-asym}).  

\bibliography{/Users/xxing/research/reference-all}

\end{document}